\documentclass[journal, twocolumn, twoside]{IEEEtran}
\usepackage{cite}
\usepackage{amsmath,amssymb,amsfonts}
\usepackage{graphicx}
\usepackage{cleveref}
\usepackage{xcolor}
\usepackage{tabularx}  %
\usepackage{lipsum}
\usepackage{mathptmx}  
\usepackage{float}
\usepackage{balance}
\usepackage{algorithm}
\usepackage[noend]{algpseudocode}
\usepackage{lipsum} 

\usepackage{booktabs}
\usepackage{adjustbox}
\usepackage{tablefootnote}
\usepackage{threeparttable} 
\usepackage{booktabs} 

\ifCLASSINFOpdf
\else
\fi

\begin{document}

\title{Voltage-Controlled Oscillator and Memristor-Based Analog Computing for Solving Systems of Linear Equations}

\author{Hao~Li, Rizwan S. Peerla, Frank~Barrows, 
Francesco~Caravelli, 
        and~Bibhu~Datta~Sahoo,~\IEEEmembership{Senior~Member,~IEEE}
\thanks{Hao Li, Rizwan S. Peerla, and~Bibhu~Datta~Sahoo are with the Department
of Electrical Engineering, University at Buffalo, Buffalo,
NY, 14260 USA (e-mail: bibhu@buffalo.edu).}
\thanks{Frank Barrows is with the Center for Nonlinear Studies and  T4 Quantum and Condensed Matter Divisions, Los Alamos National Laboratory, Los Alamos, USA (e-mail: fbarrows@lanl.gov).}
\thanks{Francesco Caravelli is with the  T4 Quantum and Condensed Matter Division, Los Alamos National Laboratory, Los Alamos, USA (e-mail: caravelli@lanl.gov).}
}

\maketitle


\begin{abstract}
Matrix computations have become increasingly significant in many data-driven applications; however, Moore's law for digital computers has been gradually approaching its limit in recent years. Moreover, digital computers encounter substantial complexity when performing matrix computations and need a long time to finish the computations, and existing analog matrix computation schemes require a large chip area and power consumption. This paper proposes a linear algebra system of equations based on integrators, which features low power consumption, compact area, and fast computation time. The demonstrated scheme is capable of performing first-order ordinary differential equation (ODE) computations to realize a 2 × 2 computation matrix and can be expanded further depending on the available chip area. Due to the simple structure of the ring oscillator, the ring oscillator-based integrator exhibits a compact area and low power consumption. Therefore, ring oscillator-based integrators are introduced into the linear algebra system of equations, and this system can be used to compute the linear algebra equations of the matrix with either positive or negative values. This paper provides a detailed analysis and verification of the proposed circuit structure. Compared to similar circuits, this work has significant advantages in terms of area, power consumption, and computation speed.

\end{abstract}

\begin{IEEEkeywords}
Matrix computation, linear algebra, ordinary differential equations, integrator, ring oscillator.
\end{IEEEkeywords}

\IEEEpeerreviewmaketitle

\section{Introduction}

\IEEEPARstart{T}{he} development of technologies such as large language models (LLM) and the Internet of Things (IoT) has led to an explosive increase in matrix computations, making the computation of digital matrices based on digital computers of the Von Neumann architecture increasingly inadequate for the demands of data processing. Analog computing typically involves simpler structures and faster computation speeds \cite{b1,b2,b3,b4,b5,b6,b7,b8,b9,b10}. Analog matrix computation methods have been revisited to overcome the bottleneck of digital matrix computation. 
The iterative process of solving linear equations using analog matrix computation can be regarded as the solution of ordinary differential equations (ODEs) \cite{b11}.  Therefore, we can transform the linear problems of analog computations into ODEs. 

There are two primary approaches to the implementation of ODEs if we use analog computation methods: one utilizes memristors in combination with operational amplifiers \cite{b12,b13,b14,b15,b16,b17,b18,b19,b20,b21}, while the other only employs analog and mixed signal circuits \cite{b22,b23,b24,b25,b26,b27}. Theoretically, analog matrix computation using memristors provides unparalleled advantages in energy efficiency and computational speed. However, its limitations include reduced precision and limited endurance for frequent write operations \cite{b28}. The advantages of using traditional analog-and-digital mixed methods (ADM) include the ability to achieve high precision and programmability, with the technology being well-established. However, it is clear that it will have higher power consumption and larger area requirements, since operational amplifiers are used to implement the integration function and variable gain amplifiers are used to implement the values of the matrix.

The traditional ADM architecture is composed of mature digital and analog circuits, utilizing well-established CMOS technology. It includes key modules such as integrators, multipliers, digital-to-analog converters (DACs), analog-to-digital converters (ADCs), SRAM, and SPI. As evidenced in \cite{b15}, integrators and multipliers are the main contributors to chip area and power consumption due to their extensive use in the design. Therefore, optimizing the design of integrators and multipliers can significantly improve the overall efficiency and performance of the chip.

This paper employs the high-precision ADM solution while utilizing alternative types of integrators and multipliers. The proposed scheme uses small-area, low-power ring oscillators as the integrator, which significantly reduces the system's area and power consumption. It achieves more efficient performance in linear algebraic operations compared to digital computation systems. An integrator based on a ring oscillator and its application in linear algebraic systems are presented. The proposed circuit implements linear algebra for ordinary differential equations and the results are analyzed and compared. In this circuit, the matrix utilized for analog computation can encompass both positive and negative values simultaneously. The approaches for generating positive and negative values of the matrix within the circuit have been systematically analyzed and validated. As resistors are employed to represent the matrix values in the proposed system, the on-chip implementation of resistors and switches incurs significant area overhead. To address this issue, memristors are utilized in place of conventional resistors, allowing each matrix value to achieve 8-bit resolution.

\section{A Linear Algebra System of Equations}
\subsection{Memristors and Linear Algebra Applications}
Beyond power consumption concerns, incorporating variable resistance or memristive devices is well motivated by iterative algorithms for solving linear algebra problems.  The inherent multiple resistance states of these devices enables direct hardware implementation of iterative algorithms. Iterative algorithms are widely used, particularly in the case of large or sparse matrices.  Iterative methods have a long history, going back to Newton's method for finding roots of functions, to Richardson iteration for general linear algebra problems, to the modern era where iterative methods, such as the Conjugate-Gradient method, are widely used for solving symmetric positive definite linear systems and linear algebra problems. Such iterative algorithms are well suited to large and sparse matrices \cite{Higham2002} and remain an essential tool for linear algebra problems that scale poorly. For example, the von Neumann-Ulam algorithm is a probabilistic method for large-scale distributed matrix inversion, by iterative refinement and random sampling, this algorithm approximates a matrix inverse \cite{forsythe1958, wu2021, sabelfeld2010}.

For these iterative methods, variable resistance states enable in memory computation, circumventing the von Neumann bottleneck. Further, the inherent dynamics of memristive devices can be leveraged to perform computations. Recent work has demonstrated an efficient and scalable iterative linear algebra solver and matrix inversion algorithm that leverages the memristive hardware dynamics \cite{LinBarrowsCarevelli2024}. 

Memristive devices are circuit elements with variable resistance; the resistance depends on the history of applied bias or current and thus functions as a memory \cite{Strukov_2008,reviewCarCar,Kumar_2022,Xiao_2023}. 
{Memristance relates the resistance of the system to the history of applied electrical stimulus.
In general, memristive behavior is a two-terminal phenomenon that leads to changes in the resistance of a material under an electrical stimulus. This change in resistance either relaxes to an initial value or is stable, corresponding to volatile and nonvolatile resistive switching.
While there is debate in the literature as to whether a system with variable resistance via the resistive switching process is a true memristor, these systems are commonly called memristors \cite{Chua-ApplPhyA-2011,Vongehr-SciRep-2015}.

Memristance is now understood to be a widespread phenomenon in nature, particularly at the nanoscale \cite{Pershin2011}.
Consequently, memristive devices have attracted significant research interest, with applications spanning scalable memory technologies, machine learning, and reservoir computing  \cite{DCRAM,Carbajal2015,reviewCarCar,reservoirmem,milano001,zhurc,Xu_FNano_2021}. Memristor crossbar arrays remain a hardware accelerator of matrix-matrix and matrix-vector multiplication \cite{mehonic}.  Previous works have generalized resistive crossbar arrays for other linear algebra problems and matrix inversion, all of which leverage the resistive memory devices (RRAM) to perform efficient matrix vector multiplication \cite{Sun2019,LinBarrowsCarevelli2024}.  These methods involve embedding in the resistive elements the variable values and iteratively obtaining desired resistance values. Here we generalize this approach, removing variable state resistors from cross bar arrays optimized for matrix vector multiplication and incorporating them in integrators. Such implementations are tailored to ODEs as it enables individual memristors to obtain iterative feedback with the analog integrator.
\vspace{-1em}

\subsection{Ring Oscillator Based Integrator}
The integrator can be realized using an operational amplifier; however, it comes at the cost of increased power consumption and large area requirements \cite{b29,b30,b31}. The ring oscillator can act as an integrator \cite{b32,b33}, as shown in Fig. 1, The main components of this integrator are a three-stage ring oscillator and three XOR-based phase detectors. The relationship between the control voltage $V_{in}$ and the oscillation frequency $\omega_{VCO}$ is given by:
\begin{equation}
    \frac{d\omega_{VCO}}{dV_{in}} = K_{VCO} 
    \label{eq:vco_relation},
\end{equation}
where $K_{VCO}$ represents the gain of the voltage-controlled oscillator (VCO). The reference phase $\varphi_{ref}$ is used to convert the output signal of the oscillator into a pulse width modulated signal (PWM). By shorting the output of the XOR gate through resistors, the PWM tones are shifted to a higher frequency. In Fig. \ref{fig:vcobg}, a three-stage ring oscillator-based integrator is presented, resulting in the PWM tones being pushed to three times the frequency of $\varphi_{ref}$. If the gain of each XOR phase detector is denoted as $K_{PD}$, the relationship between the input voltage $V_{in}$ and the output voltage $V_{out}$ can be expressed in the frequency domain as:
\begin{equation}
    \frac{V_{\text{out}}}{V_{\text{in}}} (s)= \frac{K_{\text{VCO}} K_{\text{PD}}}{s}
    \label{eq:phase_detector_output}.
\end{equation}
This integrator can be implemented when the supply voltage is less than 1V. By increasing the number of stages in the ring oscillator, the PWM tones can be pushed to higher frequencies. Compared to integrators with other architectures, the integrator offers the advantages of a smaller area and lower power consumption.
 
\begin{figure}[htbp]
\centering
\includegraphics[scale=0.8]{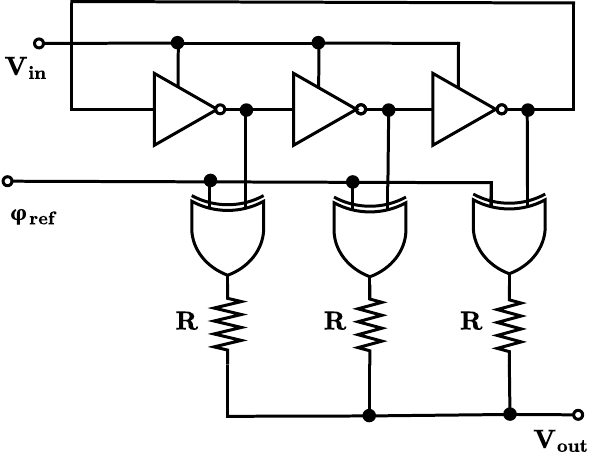}
\caption{The ring oscillator based integrator.}
\vspace{-1em}
\label{fig:vcobg}
\end{figure}

\subsection{System Architecture}
Unlike previous work, where the integrator is implemented using an operational amplifier and the multiplier is realized with a variable gain amplifier as presented in \cite{b34}, the integrator in this paper is based on a ring oscillator. As shown in Fig. 2, the feedback resistances $R_{f,00}$, $R_{f,01}$, $R_{f,10}$ and $R_{f,11}$ are placed between the input and output of the integrator, respectively, working together with the input resistances $R_{in0}$ and $R_{in1}$ to serve as the multiplier. The inputs of this system are $b_0$ and $b_1$, and can be positive or negative values; the outputs $x_0$ and $x_1$ are the solutions of the equation. The simple two-variable linear equation system formed by this circuit is:
\begin{equation}
\frac{d}{dt} 
\begin{bmatrix} x_0(t) \\ x_1(t) \end{bmatrix} 
=
\begin{bmatrix} b_0 \\ b_1 \end{bmatrix} 
- 
\begin{bmatrix} a_{00} & a_{01} \\ a_{10} & a_{11} \end{bmatrix} 
\begin{bmatrix} x_0(t) \\ x_1(t) \end{bmatrix},
\label{eq:lineareq}
\end{equation}
where we have:
\begin{equation}
\frac{d}{dt} 
\begin{bmatrix} x_0(t) \\ x_1(t) \end{bmatrix} 
=
\begin{bmatrix} u_0 \\ u_1 \end{bmatrix} ,
\label{eq:placeholder_label}
\end{equation}
and $u_0$ and $u_1$ will be close to zero as the output gradually converges. Therefore, the linear algebraic equation solved by the 2 × 1 input system is: 
\begin{equation}
 \begin{bmatrix} a_{00} & a_{01} \\ a_{10} & a_{11} \end{bmatrix}\begin{bmatrix} x_{0} \\ x_{1}  \end{bmatrix}=\begin{bmatrix} b_{0} \\ b_{1}  \end{bmatrix},
\label{eq:linear2x1}
\end{equation}
where we have:
\begin{equation}
    a_{00} = -\frac{R_{in0}}{R_{f,00}}; ~ a_{01} = -\frac{R_{in0}}{R_{f,01}} ;~ \\  
    a_{10} = -\frac{R_{in1}}{R_{f,10}}; ~a_{11} = -\frac{R_{in1}}{R_{f,11}} \\ ,
\label{eq:a}
\end{equation}
from (6), it can be seen that in this system, we can adjust the values of $a_{00}$, $a_{01}$, $a_{10}$, and $a_{11}$ by varying these resistances. It should be noted that the system in Fig. \ref{fig:sys2} can only generate a matrix with negative values. If we consider only the feedback path of $a_{00}$ as the subject of study, we can obtain the transfer function:
\begin{equation}
    H(s)= \frac{x_{\text{0}}}{b_{\text{0}}} (s)= -\frac{R_{\text{f,00}}}{R_{\text{in0}}}\frac{1}{1+\frac{s}{K_{\text{VCO}} K_{\text{PD}}}(1+\frac{R_{\text{f,00}}}{R_{\text{in0}}})}
    \label{eq:phase_detector_output},
\end{equation}
which results in a 3-dB bandwidth:
\begin{equation}
    BW= \frac{K_{\text{VCO}} K_{\text{PD}}}{1+\frac{R_{\text{f,00}}}{R_{\text{in0}}}}
    \label{eq:phase_detector_output1}.
\end{equation}

Therefore, the bandwidth can be changed by changing $K_{VCO}$, $K_{PD}$, $R_{f,00}$ and $R_{in0}$. Since $K_{PD}$ is a fixed value, if we increase $K_{VCO}$, the bandwidth will increase, so the frequency of the circuit can be adjusted faster and the converging speed will also increase efficiently. The entire system consists of two integrators based on the ring oscillator and six resistors, achieving low power consumption, a wide bandwidth, and a compact size.
\begin{figure}[htbp]
\centering
\includegraphics[scale=0.8]{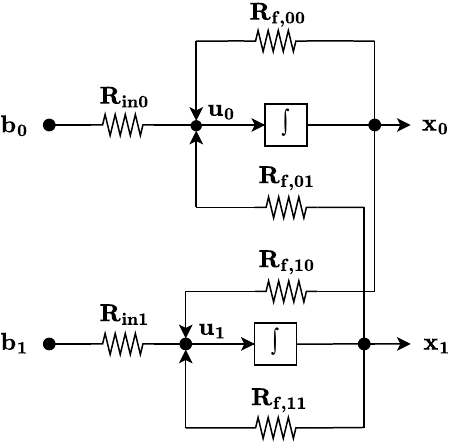}
\caption{A system of linear algebra equations having a 2 × 2 negative matrix values.}
\vspace{-1em}
\label{fig:sys2}
\end{figure}

Furthermore, generating positive values in the matrix is also feasible. As shown in Fig. \ref{fig:sys2p}, as an example, we consider the implementation in which $a_{00}$ takes a positive value. The integrator-based inverter is introduced in the feedback path, and this integrator is also realized by a ring oscillator. In this system, the generated matrix values are respectively given as follows:
\begin{equation}
    a_{00} = \frac{R_{in0}}{R_{f,00}}; ~ a_{01} = \frac{R_{in0}}{R_{f,01}} ;~ \\  
    a_{10} = \frac{R_{in1}}{R_{f,10}}; ~a_{11} = \frac{R_{in1}}{R_{f,11}} \\ ,
\label{eq:a1}
\end{equation}
where these positive values can be adjusted by changing the values of resistances. Compared to the previous system that can generate negative values, this system will utilize four additional integrators and eight resistances.
\begin{figure}[htbp]
\centering
\includegraphics[scale=0.8]{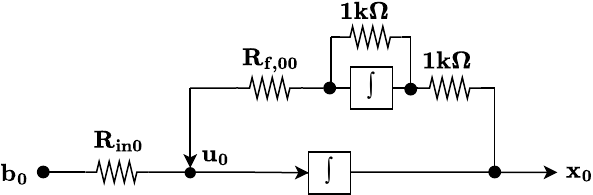}
\caption{The implementation of positive matrix values.}
\label{fig:sys2p}
\end{figure}

It is obvious that by combining the two systems that can generate positive and negative values, a matrix containing both positive and negative values can be obtained. As shown in Fig. \ref{fig:sys2m}, by combining the two systems, each matrix element can be formed as a positive or negative value. Depending on the states of the switches $S_-$ and $S_+$, each matrix element can be positive or negative. Therefore, the numerical values of the matrix implemented by this system are:
\begin{align}
a_{00} &= -\frac{R_{in0}}{R_{f,00-}} \quad \text{or} \quad \frac{R_{in0}}{R_{f,00+}} ; \\
a_{01} &= -\frac{R_{in0}}{R_{f,01-}} \quad \text{or} \quad \frac{R_{in0}}{R_{f,01+}} ; \\
a_{10} &= -\frac{R_{in1}}{R_{f,10-}} \quad \text{or} \quad \frac{R_{in1}}{R_{f,10+}} ; \\
a_{11} &= -\frac{R_{in1}}{R_{f,11-}} \quad \text{or} \quad \frac{R_{in1}}{R_{f,11+}},
\label{eq:placeholder_label}
\end{align}
where we can see that each value is determined by the resistance values of two resistors when $R_{in0}$ and $R_{in1}$ are fixed at specific values. We can obtain the desired matrix based on the numerical relationships in (10), (11), (12), and (13).
\begin{figure}[htbp]
\centering
\includegraphics[scale=0.8]{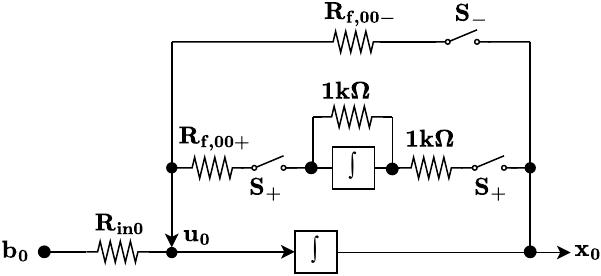}
\caption{The implementation of mixed matrix values.}
\label{fig:sys2m}
\end{figure}

When implementing the matrix values using the aforementioned method, a higher proportion of positive elements in the matrix will lead to an increased requirement for integrators. In addition to the integrators used on the main signal path, the number of additional integrators required is equal to the number of positive values in the matrix. To reduce the number of integrators, the integrator-based inverter can be reused. Specifically, when the number of positive elements in the matrix is less than the number of negative elements, inverters are used to realize the positive elements. Otherwise, the matrix is negated so that the originally negative elements become positive, which are then realized by inverters. Considering the implementation of a $N$~×~$N$ matrix, Table \ref{tab:vco} summarizes the comparison of the number of integrators required before and after the application of the proposed method. As shown in this table, the reuse of inverters leads to a reduction in the number of integrators, which is given by:
\begin{equation*}
N^2 - \left\lfloor \dfrac{N^2}{2} \right\rfloor.
\end{equation*}
As an example, consider an $8$~×~$8$ matrix, where this method results in a saving of 32 integrators. In MIMO systems, the covariance matrix is typically symmetric and positive semi-definite. This property enables a reduction in implementation complexity, especially in analog or hardware-based systems that accumulate or store covariance values. For a full \( N \times N \) matrix, symmetry allows for storing only the upper (or lower) triangular part, reducing the number of required integrators to
\begin{equation*}
\left\lfloor \dfrac{N^2+N}{4} \right\rfloor+N,
\end{equation*}
which is significantly more efficient. The positive semi-definite property further ensures all eigenvalues are non-negative, maintaining stability in estimation and detection tasks.
\vspace{-1em}
\begin{table}[t]
\caption{The comparison of the number of integrators required}
\centering
\begin{tabular}{lllll}
\cline{1-4}
\multicolumn{1}{|l|}{}                      & \multicolumn{1}{c|}{Before Application} & \multicolumn{1}{c|}{After Application} & \multicolumn{1}{c|}{MIMO$\dag$\tnote{a}} &  \\ \cline{1-4}
\multicolumn{1}{|c|}{Integrators} & \multicolumn{1}{c|}{$N^2 + N$}             & \multicolumn{1}{c|}{$\left\lfloor \frac{N^2}{2}\ \right\rfloor + N$}    & \multicolumn{1}{c|}{$\left\lfloor \frac{N^2+N}{4}\ \right\rfloor + N$}                        &  \\ \cline{1-4}
                                            &                                         &                                        &                                               &  \\
                                            &                                         &                                        &                                               & 
\end{tabular}
\vspace{-2em}
\label{tab:vco}
\begin{tablenotes}[flushleft]
\item[a] $\dag$After application.
\vspace{-1em}
\end{tablenotes}
\end{table}

\section{Circuit Design}
\subsection{Integrator}
\indent The $G_m$-C integrator and the Opamp-RC integrator are the two most widely adopted analog integration architectures\cite{b40}. $G_m$-C integrators, which utilize operational transconductance amplifiers (OTAs) in open-loop configurations, offer bandwidths comparable to the OTAs themselves but exhibit poor linearity. While resistive degeneration can enhance linearity to some extent, its effectiveness is constrained in deep sub-micron technologies due to limitations in voltage headroom and increased thermal noise. In contrast, Opamp-RC integrators employ feedback to mitigate OTA nonlinearity, resulting in improved linearity at the cost of significantly reduced bandwidth. Both integrator types suffer from finite OTA gain, which introduces integration loss. Furthermore, as channel lengths scale down, the intrinsic transistor gain ($g_{m}r_{o}$) diminishes, exacerbating the DC gain limitations in traditional integrator designs\cite{b40}. In light of the limitations associated with OTA-based integrators, this work explores the use of CMOS inverter-based ring oscillators as an alternative for realizing integrators. These structures inherently provide infinite DC gain, which remains unaffected by variations in transistor dimensions or supply voltage.\\
\indent Consider a voltage-controlled oscillator (VCO) implemented using a ring oscillator (RO). The input control voltage ($V_{IN}$) generates output oscillation frequency ($\omega_{VCO}$) having a VCO gain as in \eqref{eq:vco_relation}:
This linear approximation is valid over the tuning range of the VCO. Because phase is the time integral of frequency, the VCO can be modeled as an ideal integrator with $V_{IN}$ and output phase $\phi(out)$. The resulting transfer function is given by \eqref{eq:kvco1}:
\begin{equation}
\frac{\phi_{out}(s)}{v_{in}(s)} = \frac{K_{VCO}}{s}
\label{eq:kvco1}
\end{equation}
\indent This relationship illustrates that the VCO acts as an ideal integrator of the input voltage, independent of transistor dimensions or supply voltage variations. This inherent property of ring oscillators enables alternative implementations of analog signal processing building blocks—such as filters—in low-voltage, deep submicron CMOS technologies. Since the ring oscillator functions as a lossless integrator only when observed in the phase domain, a phase-to-voltage or phase-to-current interface is required to connect the RO-based integrator with other circuit components. A phase detector (PD) serves this role by converting the oscillator’s phase output into a pulse-width modulated (PWM) signal. The PD determines the phase difference between the VCO output and a reference clock, generating a PWM signal whose duty cycle encodes the phase error.
\subsection{Ring Oscillator Based Integrator Implementation}
\indent The output of the integrator ($V_{OUT}$) is a PWM signal that inherently contains spurious tones at the modulation frequency and its harmonics. Recovering the underlying analog signal requires low-pass filtering, which can increase power consumption and introduce non-linearity. To mitigate this tradeoff, a multi-phase PWM architecture is adopted, wherein multiple phase-shifted PWM signals are summed to form the integrator output\cite{b40}. With $M$ phases, the transition density increases by a factor of $M$, and the combined output assumes $(M+1)$ discrete voltage levels. This enhancement in both time and amplitude resolution significantly attenuates the residual PWM tones, reducing filtering requirements and improving overall system performance. The block diagram of the 32-phase implementation of the integrator is shown in Fig. \ref{fig:basicint}. \\
\begin{figure}[!h]
\centering
\includegraphics[width=\linewidth]{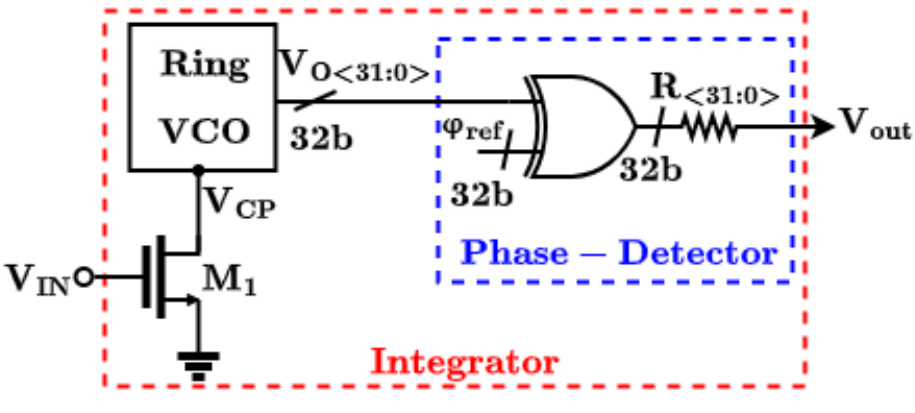}
\caption{Block diagram of single-ended RO-based integrator implementation.}
\vspace{-1em}
\label{fig:basicint}
\end{figure}
\indent The proposed RO-based integrator can be realized using three distinct architectures. Each approach utilizes a $16$-stage differential delay cell–based ring oscillator, producing $32$ phase outputs. In the first method (see Fig. \ref{fig:method1}), the outputs of all delay stages are directly connected to level shifters to achieve rail-to-rail voltage swings, which are then fed into a phase detector. Although this approach preserves phase resolution, it is power-intensive because of the large number of level shifters.\\
\begin{figure}[!h]
\centering
\includegraphics[width=\linewidth]{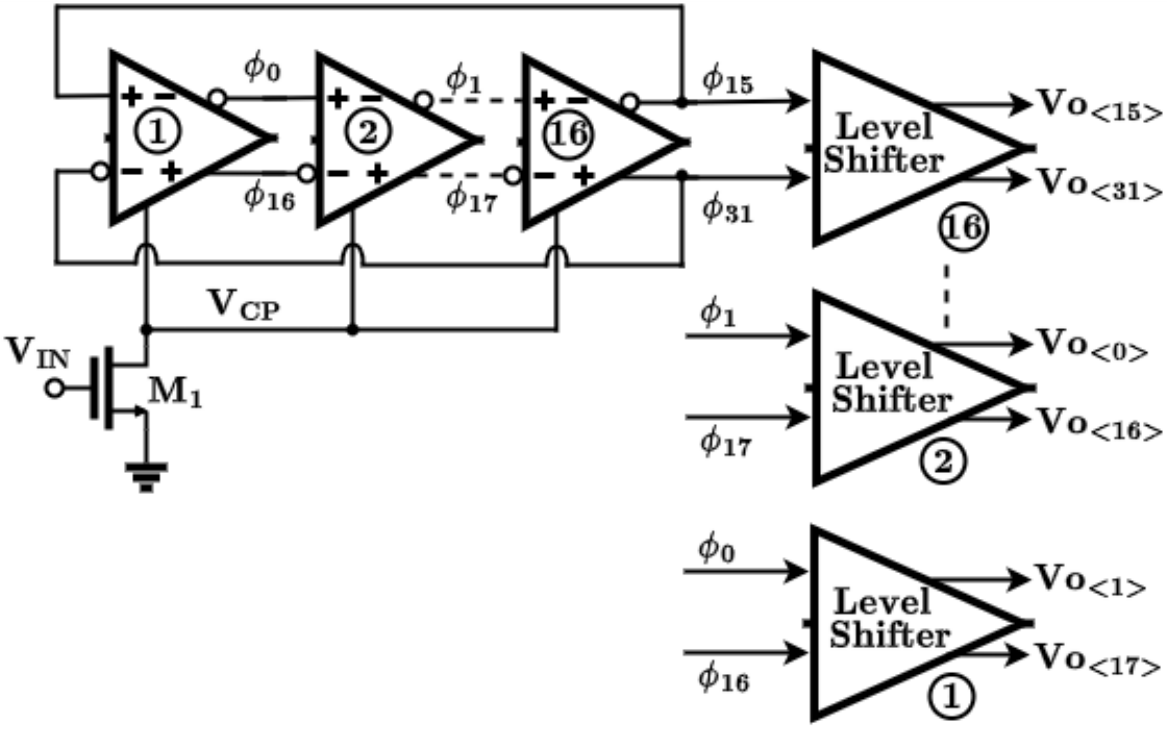}
\caption{Ring oscillator implementation using delay cells and $16$-level shifters.}
\label{fig:method1}
\end{figure}
\indent The second method (see Fig. \ref{fig:method2}) reduces power by passing only the final RO output through a single level shifter. This signal then drives a $16$-stage Johnson counter, as described in \cite{b41}, which generates $32$ peak-to-peak phase outputs. However, this method effectively divides the oscillator frequency by $16$, resulting in a significant reduction in the $K_{VCO}$. \\
\begin{figure}[!h]
\centering
\includegraphics[width=\linewidth]{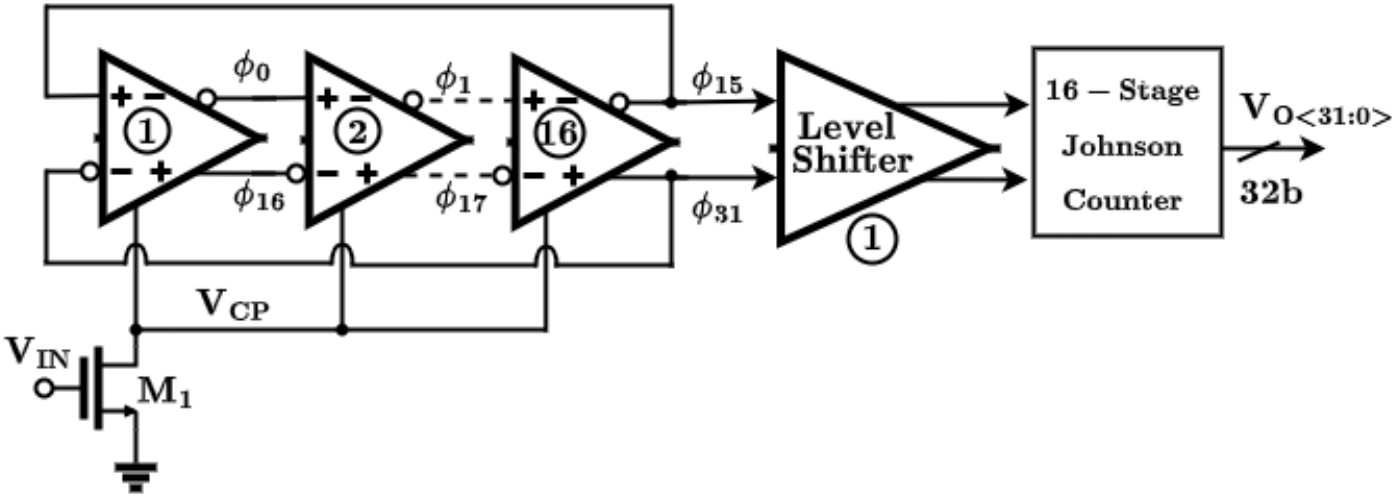}
\caption{Ring oscillator implementation using delay cells, $1$-level shifter, and $16$-stage Johnson counter.}
\label{fig:method2}
\end{figure}
\indent The third method strikes a balance between power and phase resolution (see Fig. \ref{fig:method3}). In this approach, outputs from stages $4$, $8$, $12$, and $16$ are passed through level shifters and then processed by four separate $4$-stage Johnson counters. This configuration yields $32$ output phases with moderate power consumption and $K_{VCO}$ (RO output is divided by $4$).
\begin{figure}[!h]
\centering
\includegraphics[width=\linewidth]{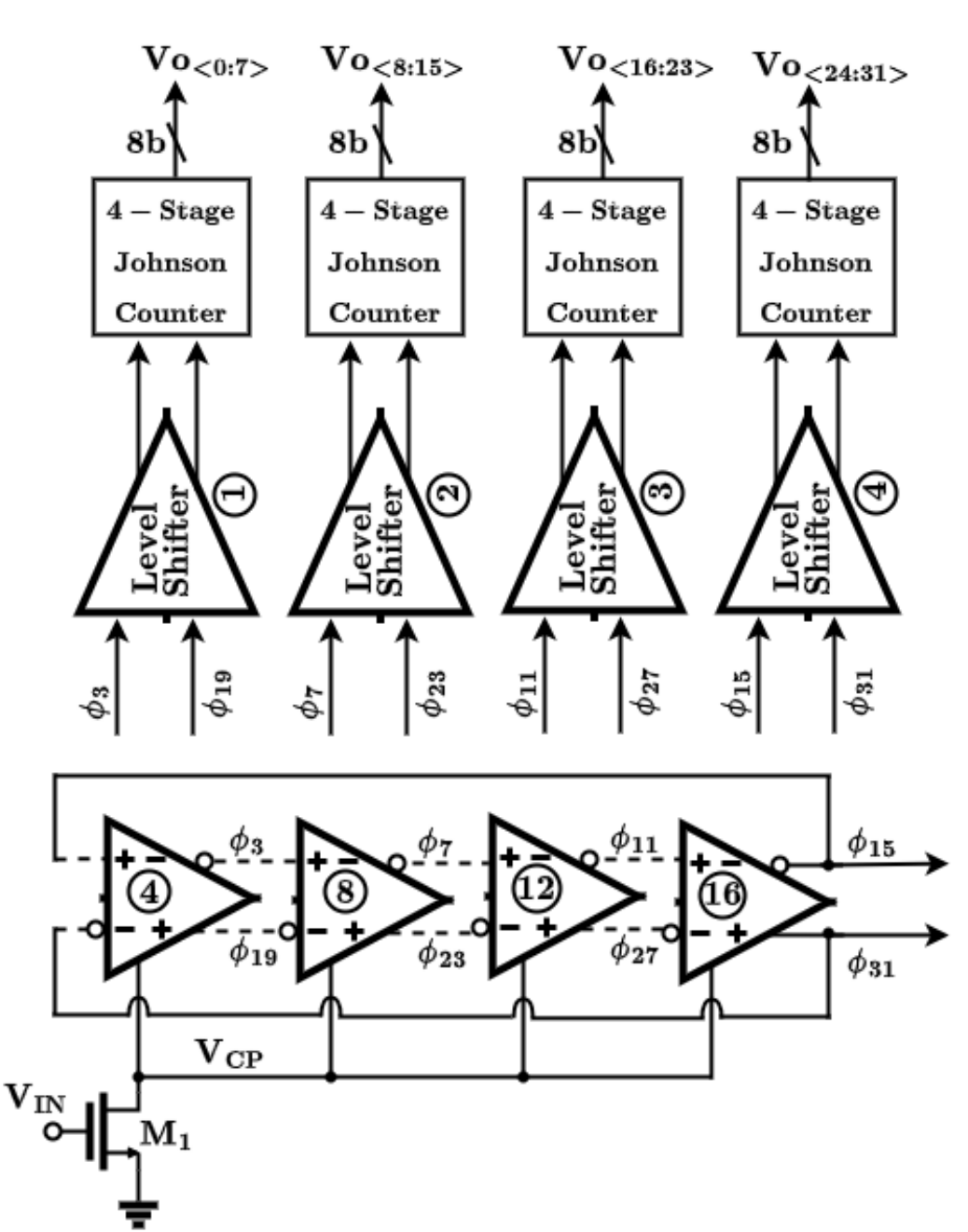}
\caption{Ring oscillator implementation using delay cells, $4$-level shifters, and four $4$-stage Johnson counters.}
\vspace{-1.5em}
\label{fig:method3}
\end{figure}

Figure \ref{fig:cco}(a) illustrates the schematic of the ring oscillator delay cell \cite{b40}. The design is realized using a pair of current-starved CMOS inverters. Feed-forward resistors are used to couple the outputs, facilitating differential operation. The number of stages was selected to define the nominal gain of the VCO. Frequency tuning is achieved by adjusting the current through the n-MOS control transistor.\\
\begin{figure}[!h]
\centering
\includegraphics[width=\linewidth]{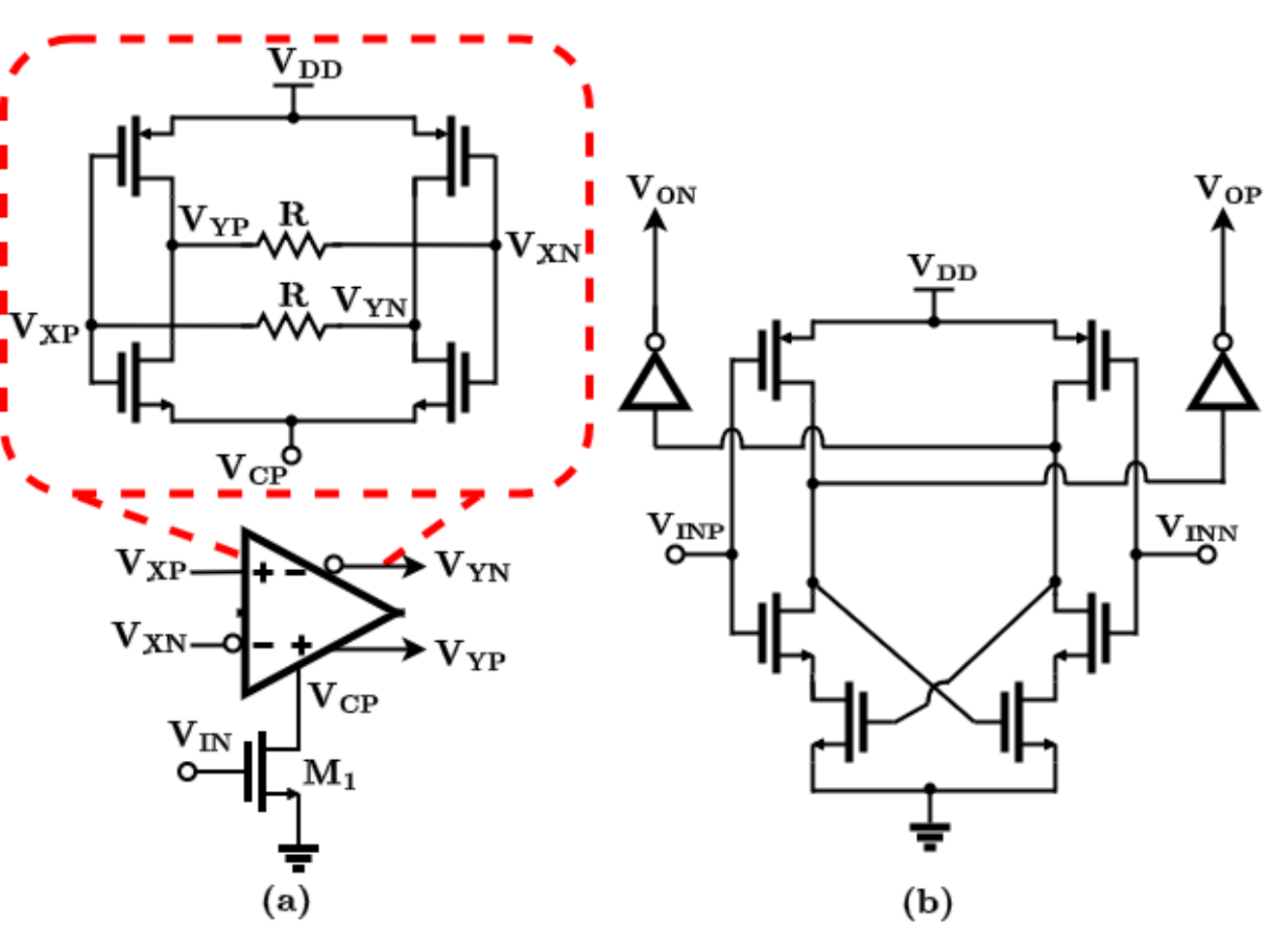}
\caption{Schematic of (a) delay cell of the ring oscillator and (b) level shifter design.}
\vspace{-1em}
\label{fig:cco}
\end{figure}
The level shifter utilized in this design, shown in Fig. \ref{fig:cco}(b), incorporates a cross-coupled latch to ensure rail-to-rail output swings from the delay cell \cite{b40}. Since the oscillator operates between $V_{INP}$ and $V_{DD}$, the oscillator phases are supply-referenced. Consequently, the level shifter employs p-MOS input devices and an n-MOS-based latch. To enhance performance, an additional n-MOS transistor is placed in series with the latch transistors, with its gate driven by the oscillator phases. This configuration enables low-delay operation without incurring static power consumption. Although the level shifter occupies minimal silicon area relative to the delay cell, it requires complementary phase signals, which are readily available from the differential oscillator.

In Fig. \ref{fig:jc}, a $4$-stage Johnson counter is shown to generate multiple clock phases \cite{b41}. It consists of a chain of four D-flip-flops (DFFs) configured similarly to a shift register, all driven by the master clock. The output of each flip-flop is connected to the input of the next, forming a sequential logic chain. Each DFF produces a signal with a frequency of ${f_{CLK}}/{4}$. Using both the true and complementary outputs of the four DFFs, eight evenly spaced clock phases are obtained, each separated by $45^{\circ}$.
\begin{figure}[!h]
\centering
\includegraphics[width=\linewidth]{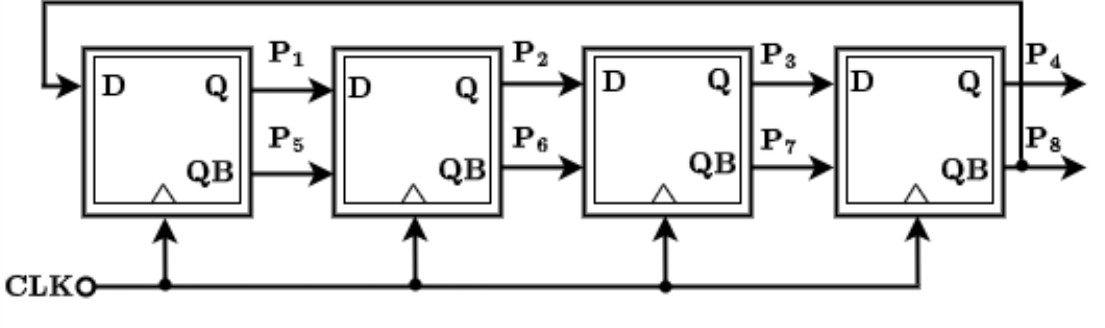}
\caption{Block diagram of $4-$stage Johnson counter that produce $8-$ phases.}
\label{fig:jc}
\end{figure}

A simulation is performed on the output of the oscillator when we use it as a low pass filter. The resulting variation of output frequency as a function of the input voltage is illustrated in Fig. \ref{fig:kvco}. The simulation results indicate that the output frequency of the oscillator demonstrates excellent linearity. Within the input voltage range of 0.5~V to 1~V, the VCO gain ($K_{VCO}$) remains approximately constant.
\begin{figure}[htbp]
\centering
\includegraphics[scale=0.2]{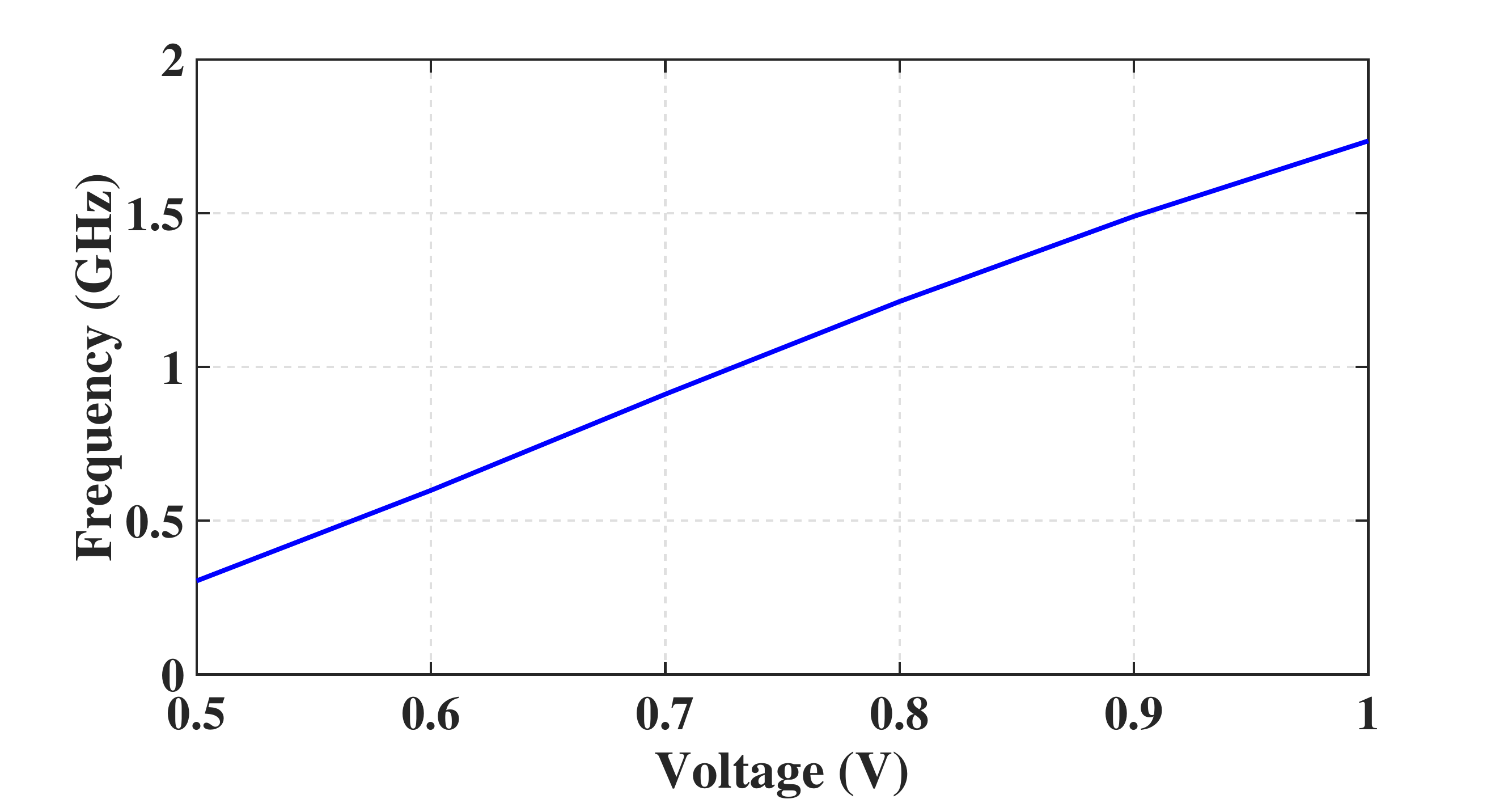}
\caption{The simulation of the relationship between the VCO input voltage and output frequency.}
\label{fig:kvco}
\end{figure}

The designed low-pass filter is simulated under a sampling frequency of 100~MHz, as shown in Fig. \ref{fig:sfdrs}. The simulation results demonstrate that the filter achieves a spurious-free dynamic range (SFDR) of 68.8~dB, indicating excellent linearity performance. 
\begin{figure}[htbp]
\centering
\includegraphics[scale=0.2]{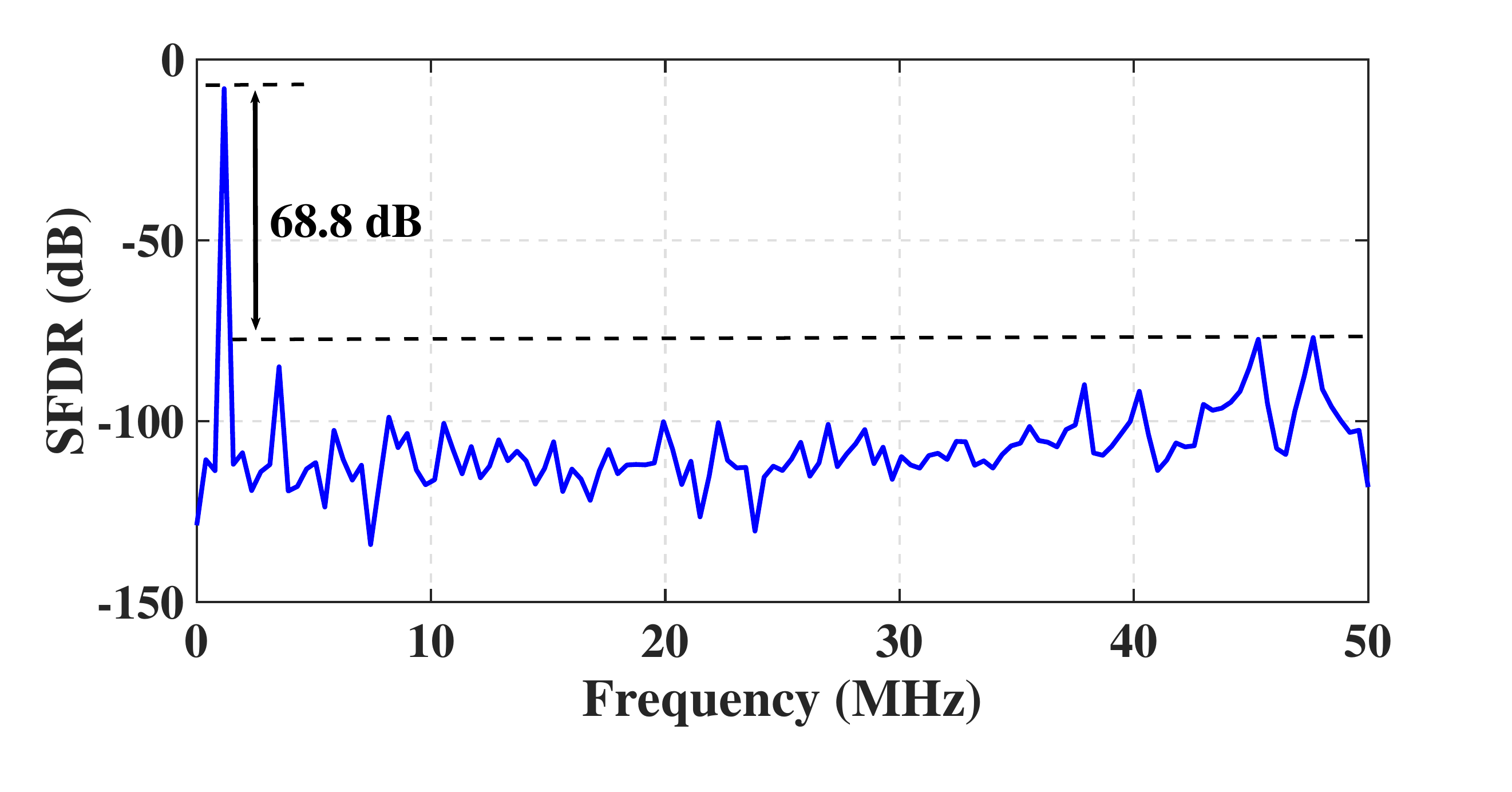}
\caption{The simulation result of the SFDR for the low-pass filter circuit.}
\label{fig:sfdrs}
\end{figure}
To evaluate linearity performance, simulations are performed on 2 × 2 and 4 × 4 matrix circuits constructed using the low-pass filter design. The output signals exhibit SFDRs of 61.8~dB and 62.2~dB, respectively, as shown in Fig.~\ref{fig:sfdr2} and Fig.~\ref{fig:sfdr4}, which demonstrate sufficient linearity for performing matrix operations.
\begin{figure}[htbp]
\centering
\includegraphics[scale=0.2]{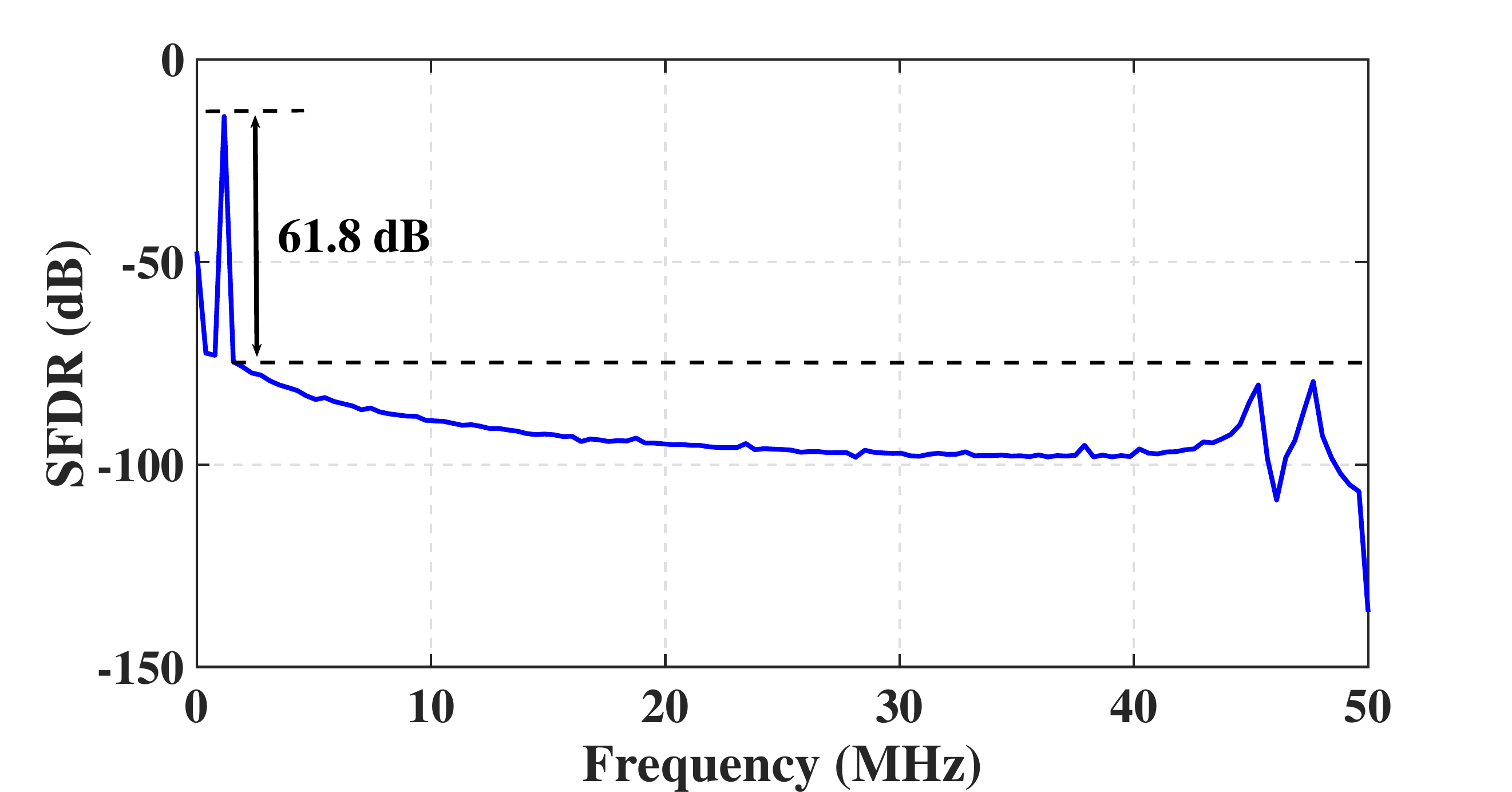}
\caption{The simulation result of the SFDR for the 2 × 2 matrix circuit.}
\vspace{-1em}
\label{fig:sfdr2}
\end{figure}
\begin{figure}[htbp]
\centering
\includegraphics[scale=0.2]{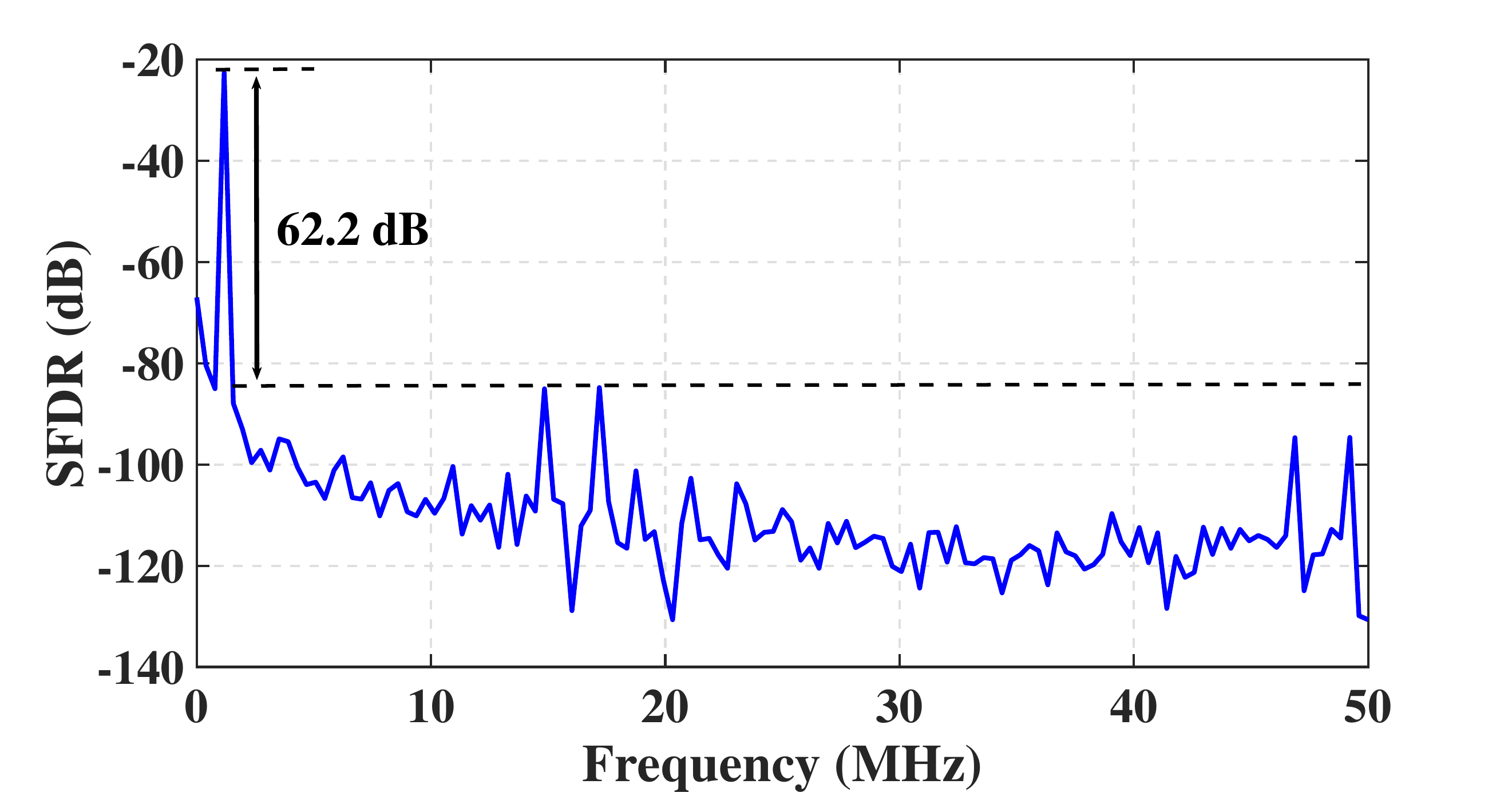}
\caption{The simulation result of the SFDR for the 4 × 4 matrix circuit.}
\vspace{-1em}
\label{fig:sfdr4}
\end{figure}
\subsection{Implementation of Feedback Resistor}
\subsubsection{Poly Resistor}
As shown in Fig. \ref{fig:program}, a 3-bit example is used to demonstrate the implementation of programmability within the circuit. In this case, the switch $S_x$ is assumed to be closed by default. When the three bits are enabled, each bit corresponds to the closure of switches $S_1$, $S_2$, and $S_3$, respectively. 
\begin{figure}[htbp]
\centering
\includegraphics[scale=0.7]{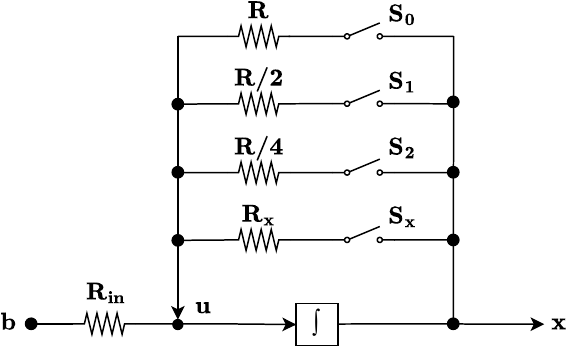}
\caption{The circuit for the 3-bit example of the programmable matrix.}
\label{fig:program}
\end{figure}
Table \ref{tab:res} summarizes the relationship between the enabled or disabled status of specific bits and the corresponding switch configurations and matrix values.
\begin{table}[t]
\caption{The relationship between the bit status and matrix values}
\centering
\resizebox{\linewidth}{!}{%
\begin{tabular}{|c|c|c|} 
\hline 
Specific bits status& Switch status& Matrix values\\ 
\hline 
000& on: $S_x$; off: $S_2$, $S_1$, $S_0$& $-R_{in}/R$\\ 
\hline 
001& on: $S_x$, $S_0$; off: $S_2$, $S_1$&  $-2R_{in}/R$\\ 
\hline
 010& on: $S_x$, $S_1$; off: $S_2$, $S_0$&  $-3R_{in}/R$\\\hline
 011& on: $S_x$, $S_0$, $S_1$; off: $S_2$& $-4R_{in}/R$\\\hline
 100& on: $S_x$, $S_2$; off: $S_1$, $S_0$&$-5R_{in}/R$\\\hline
 101& on: $S_x$, $S_2$, $S_0$; off: $S_1$&$-6R_{in}/R$\\\hline
 110& on: $S_x$, $S_2$, $S_1$; off: $S_0$&$-7R_{in}/R$\\\hline
 111& on: $S_x$, $S_2$, $S_1$, $S_0$&$-8R_{in}/R$\\\hline
\end{tabular}%
}
\label{tab:res}
\end{table}

The proposed structure is further extended to an 8-bit implementation, with individual bit-wise enable control. An input voltage of 0.25~V is applied, and with $R_{in}$ and $R$ set to 2000 $\Omega$ and 1000 $\Omega$, respectively, the simulation results are shown in Fig. \ref{fig:program_re}. The simulation results indicate that 8-bit precision in the matrix values can be achieved, provided that the on-resistance of the switches is sufficiently low.
\vspace{-1em}
\begin{figure}[htbp]
\centering
\includegraphics[scale=0.2]{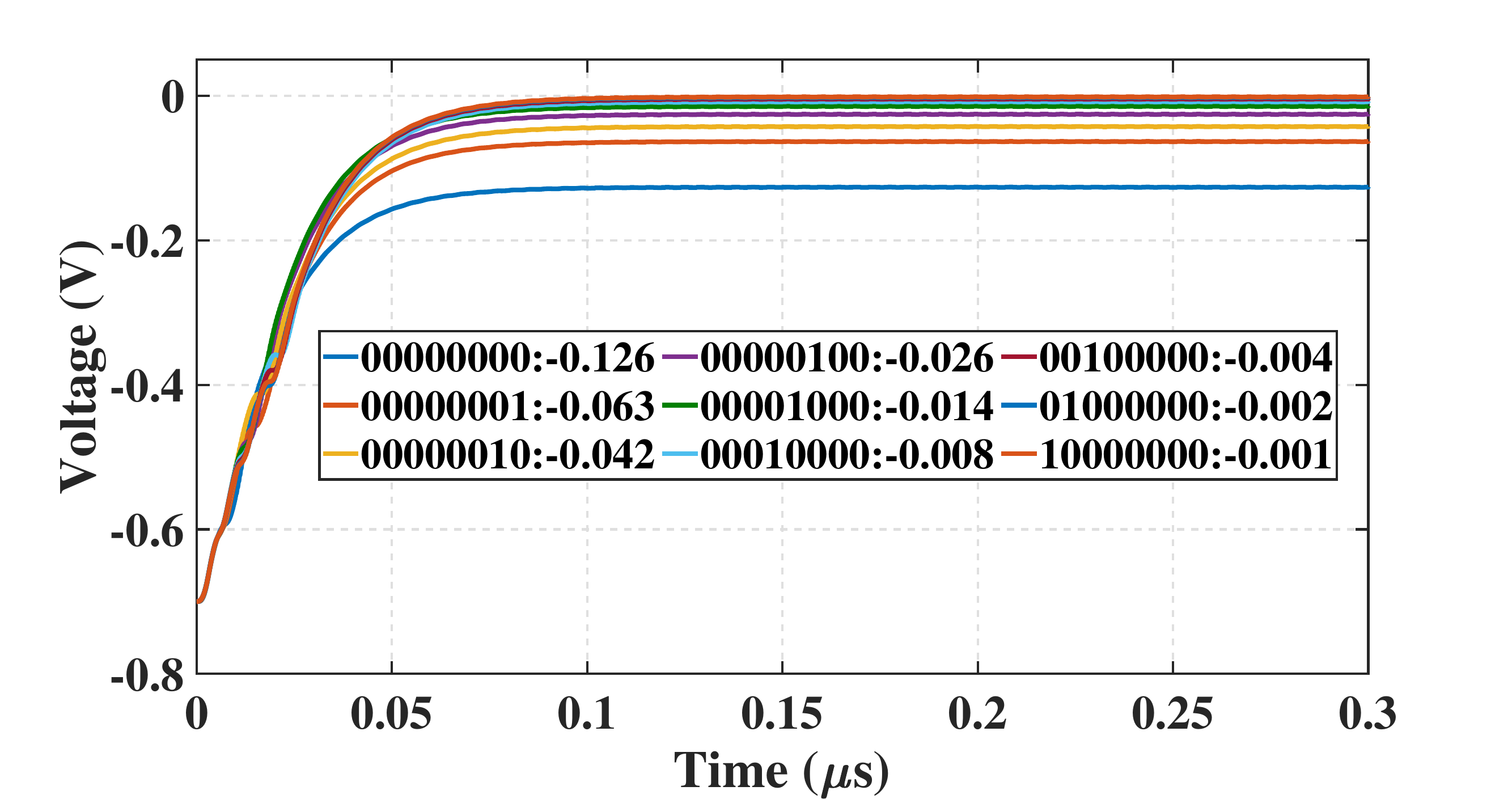}
\caption{The simulation results for the 8-bit example of the programmable matrix.}
\label{fig:program_re}
\end{figure}
\subsubsection{Memristor}
Although this configuration enables programmability, attaining low switch resistance typically requires an increased chip area. Furthermore, the necessity of a large number of resistors introduces additional area overhead. The aforementioned issues can be effectively addressed by employing memristors with tunable resistance. 

By leveraging their programmable conductance states, memristors can replace conventional resistor-switch networks, thereby significantly reducing the required chip area and simplifying the routing complexity. In addition, their non-volatile nature enables compact and energy-efficient implementation of programmable analog matrices. In the system, the feedback resistors of the integrator are replaced with memristors, and the system is evaluated using the following equation: 
\begin{equation}
 B=\begin{bmatrix} 0.07 \\ 0.02  \end{bmatrix}.
\label{eq:bmatrix}
\end{equation}
Adjusting the values of the memristors, the following matrix is implemented: 
\begin{equation}
 A=\begin{bmatrix} -4 & -1.5 \\ -2 & -1 \end{bmatrix}.
\label{eq:amatrix}
\end{equation}
As demonstrated by the simulation results in Fig. \ref{fig:program_mem}, the equation of linear algebra that has been successfully solved is:
\begin{equation}
 \begin{bmatrix} -4 & -1.5 \\ -2 & -1 \end{bmatrix}\begin{bmatrix} -0.04 \\ 0.06  \end{bmatrix}=\begin{bmatrix} 0.07 \\ 0.02 \end{bmatrix}.
\label{eq:ax}
\end{equation}
\vspace{-1.5em}
\begin{figure}[htbp]
\centering
\includegraphics[scale=0.2]{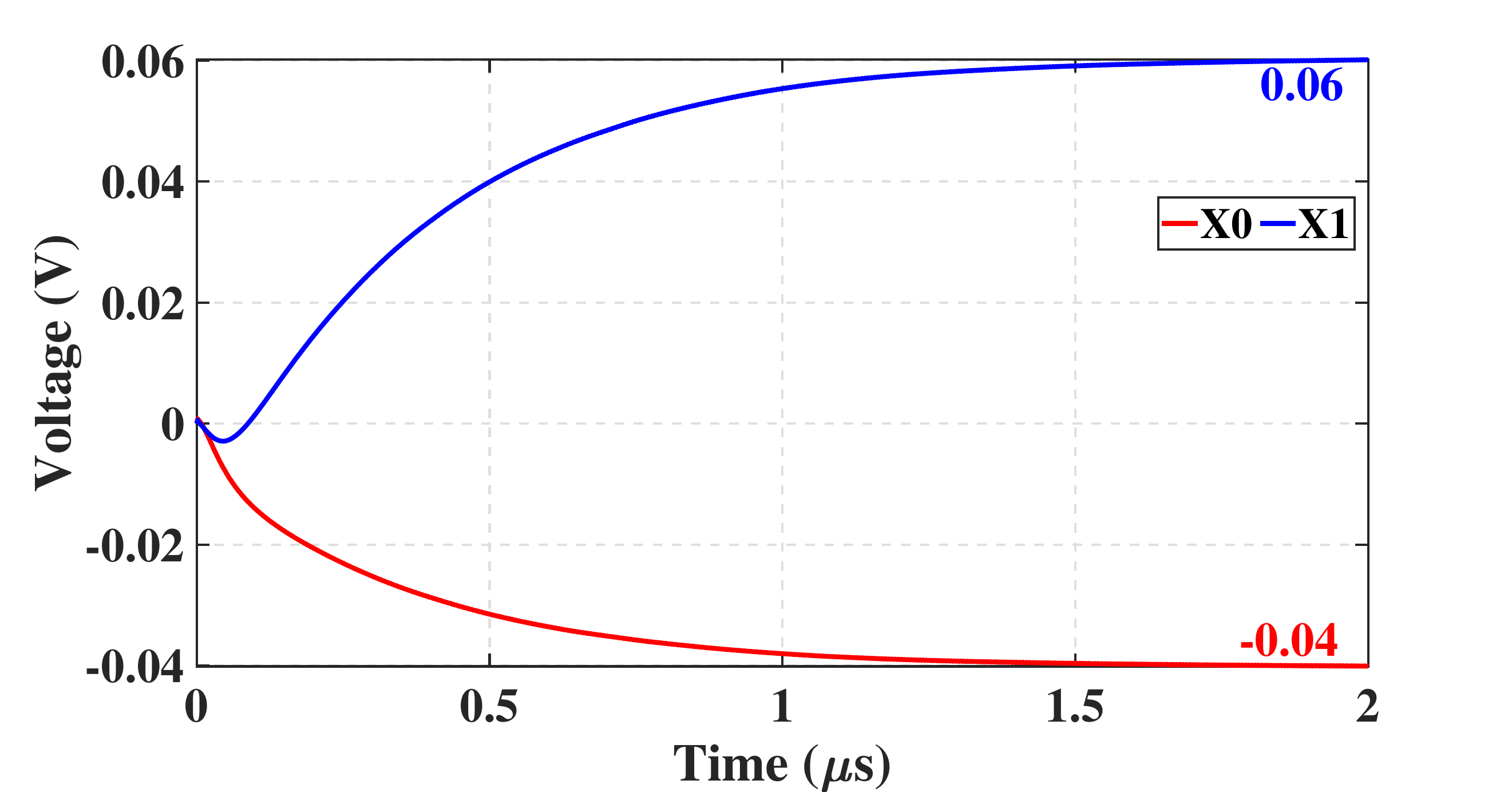}
\caption{The simulation result using memristor models.}
\vspace{-1.5em}
\label{fig:program_mem}
\end{figure}

\section{Analysis and Comparison}

\subsection{Analysis}
In Section II, we introduced the system for solving linear algebraic equations. In this subsection, we will perform an analysis and validation of the system. To solve a linear equation of the form $AX=B$, when the input $B$ of our system is a 2 × 1 matrix, according to equation (5), we set the value of $B$ to:
\begin{equation}
 B=\begin{bmatrix} 0.45 \\ 0.24  \end{bmatrix}.
\label{eq:bmatrix1}
\end{equation}
At the same time, to verify the system shown in Fig. \ref{fig:sys2}, we adjust the resistance values to set $A$ to: 
\begin{equation}
 A=\begin{bmatrix} -4 & -1.5 \\ -2 & -1 \end{bmatrix}.
 \label{eq:amatrix1}
\end{equation}
When we applying the values of $A$ to the system where  the $K_{vco}$ of the oscillator is 300 MHz, the simulation results are shown in Fig. \ref{fig:2n}, where we can get:
\begin{equation}
 X=\begin{bmatrix} -0.09 \\ -0.06  \end{bmatrix},
\label{eq:xmatrix}
\end{equation}
thus, we successfully achieve the following equation:
\begin{equation}
 \begin{bmatrix} -4 & -1.5 \\ -2 & -1 \end{bmatrix}\begin{bmatrix} -0.09 \\ -0.06  \end{bmatrix}=\begin{bmatrix} 0.45 \\ 0.24  \end{bmatrix}.
\label{eq:ax1}
\end{equation}

\begin{figure}[htbp]
\centering
\includegraphics[scale=0.2]{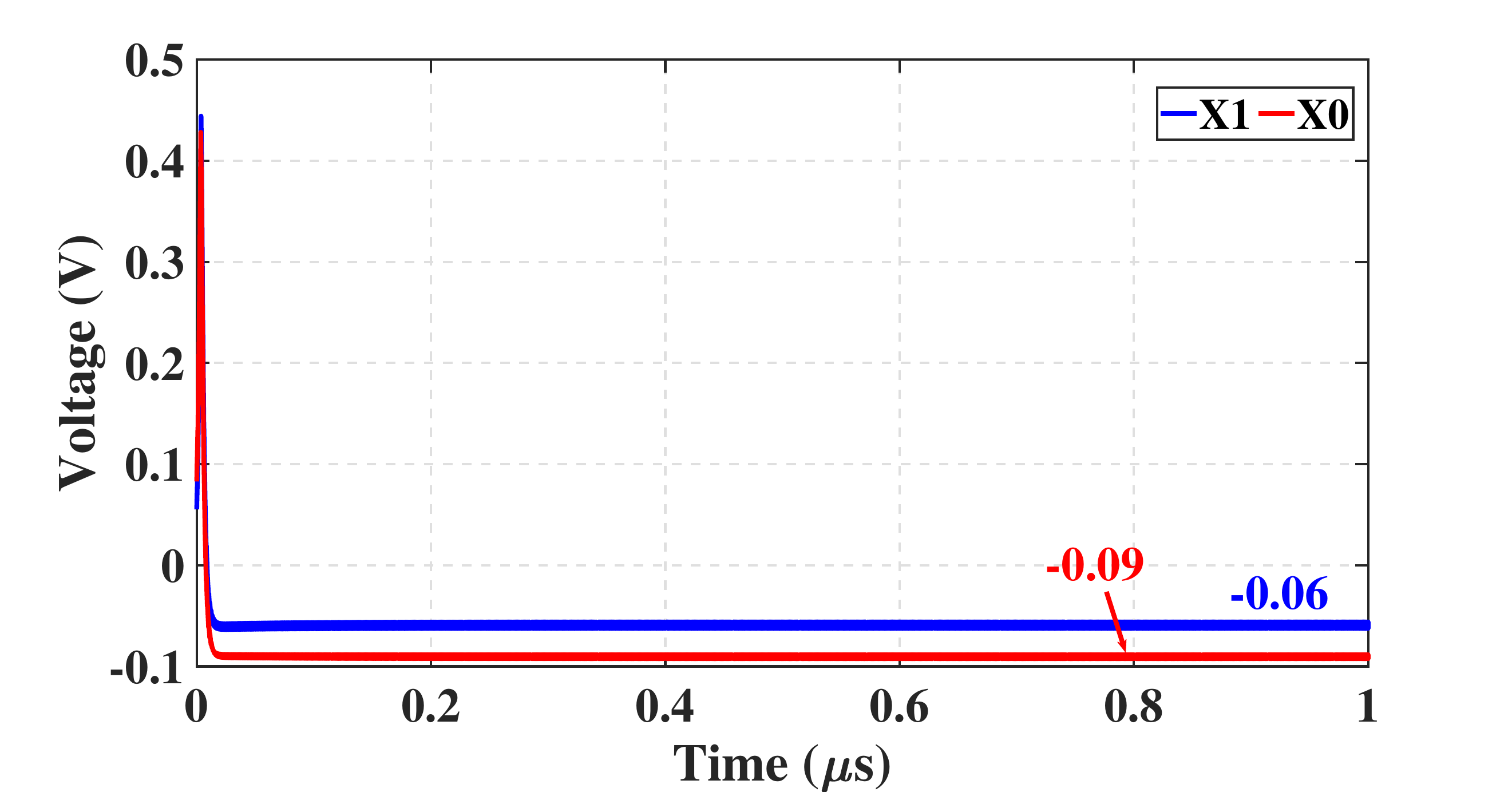}
\caption{The output of the system with a 2 × 2 negative matrix values.}
\vspace{-1em}
\label{fig:2n}
\end{figure}

In Fig. \ref{fig:sys2p}, all the matrix values in the system are positive. We keep the value of $B$ unchanged, and adjust the resistance values to ensure that:
\begin{equation}
 A=\begin{bmatrix} 4 & 1.5 \\ 2 & 1 \end{bmatrix}.
\label{eq:amtrix2}
\end{equation}
Following the system in Fig. \ref{fig:sys2p}, the output in Fig. \ref{fig:2p} is:
\begin{equation}
 X=\begin{bmatrix} 0.09 \\ 0.06  \end{bmatrix},
\label{eq:xmatrix2}
\end{equation}
and we already get the linear algebra equation of:
\begin{equation}
 \begin{bmatrix} 4 & 1.5 \\ 2 & 1 \end{bmatrix}\begin{bmatrix} 0.09 \\ 0.06  \end{bmatrix}=\begin{bmatrix} 0.45 \\ 0.24  \end{bmatrix}.
\label{eq:ax2}
\end{equation}
\vspace{-1em}
\begin{figure}[htbp]
\centering
\includegraphics[scale=0.2]{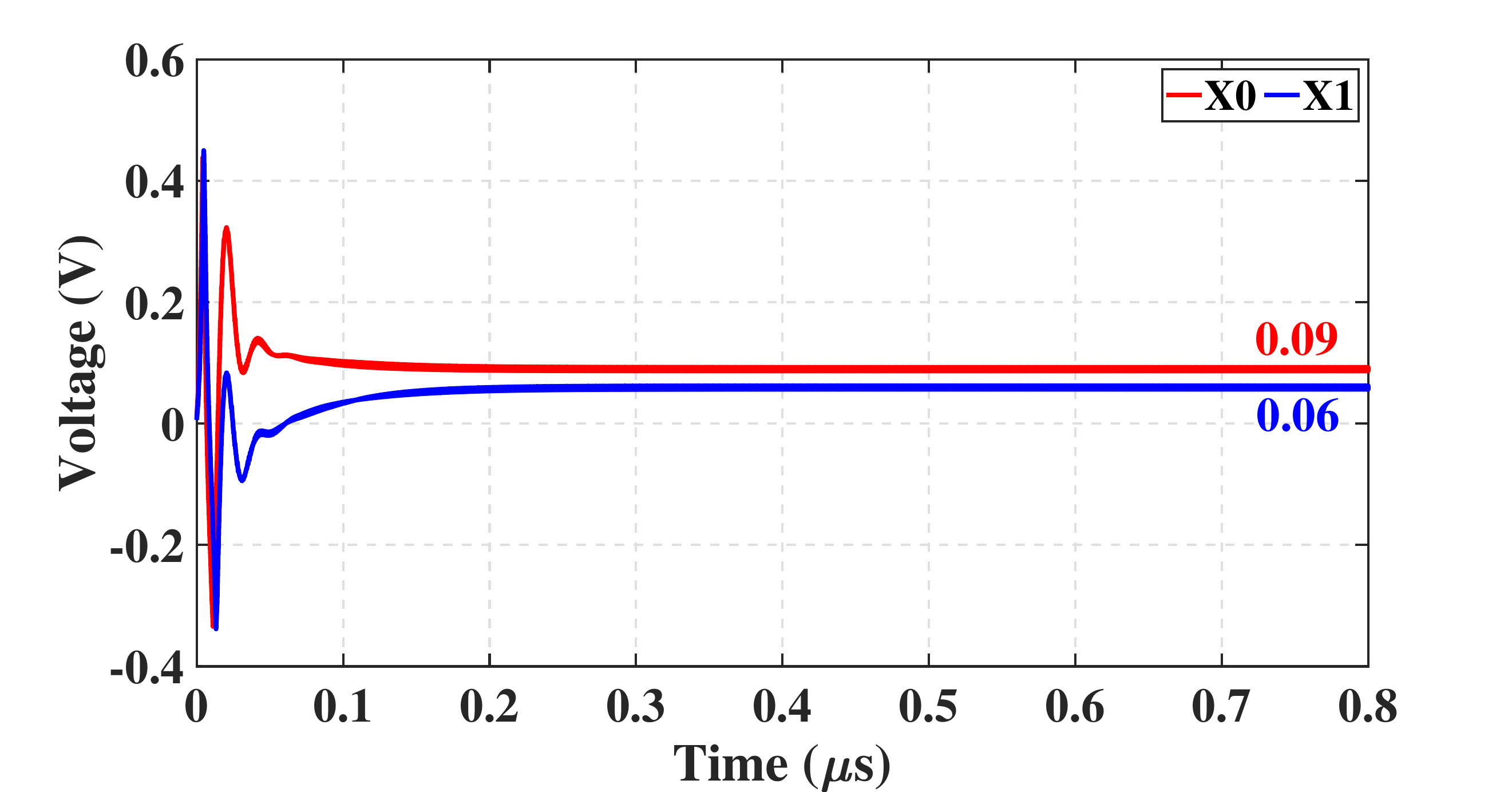}
\caption{The output of the system with a 2 × 2 positive matrix values.}
\label{fig:2p}
\end{figure}

As mentioned above, if we combine the positive and negative values, we can obtain $A$ which contains both positive and negative values. We continue to keep $B$ unchanged while adjusting the resistance to modify the value of $A$ to:
\begin{equation}
 A=\begin{bmatrix} -4 & 1.5 \\ -2 & 1 \end{bmatrix}.
\label{eq:amatrix3}
\end{equation}
The system output obtained through simulation is shown in Fig. \ref{fig:2mixed}, where we get the following output values:
\begin{equation}
 X=\begin{bmatrix} -0.09 \\ 0.06  \end{bmatrix},
\label{eq:xmatrix3}
\end{equation}
and this equation:
\begin{equation}
 \begin{bmatrix} -4 & 1.5 \\-2 & 1 \end{bmatrix}\begin{bmatrix} -0.09 \\ 0.06  \end{bmatrix}=\begin{bmatrix} 0.45 \\ 0.24  \end{bmatrix}.
\label{eq:ax3}
\end{equation}
\begin{figure}[htbp]
\centering
\includegraphics[scale=0.2]{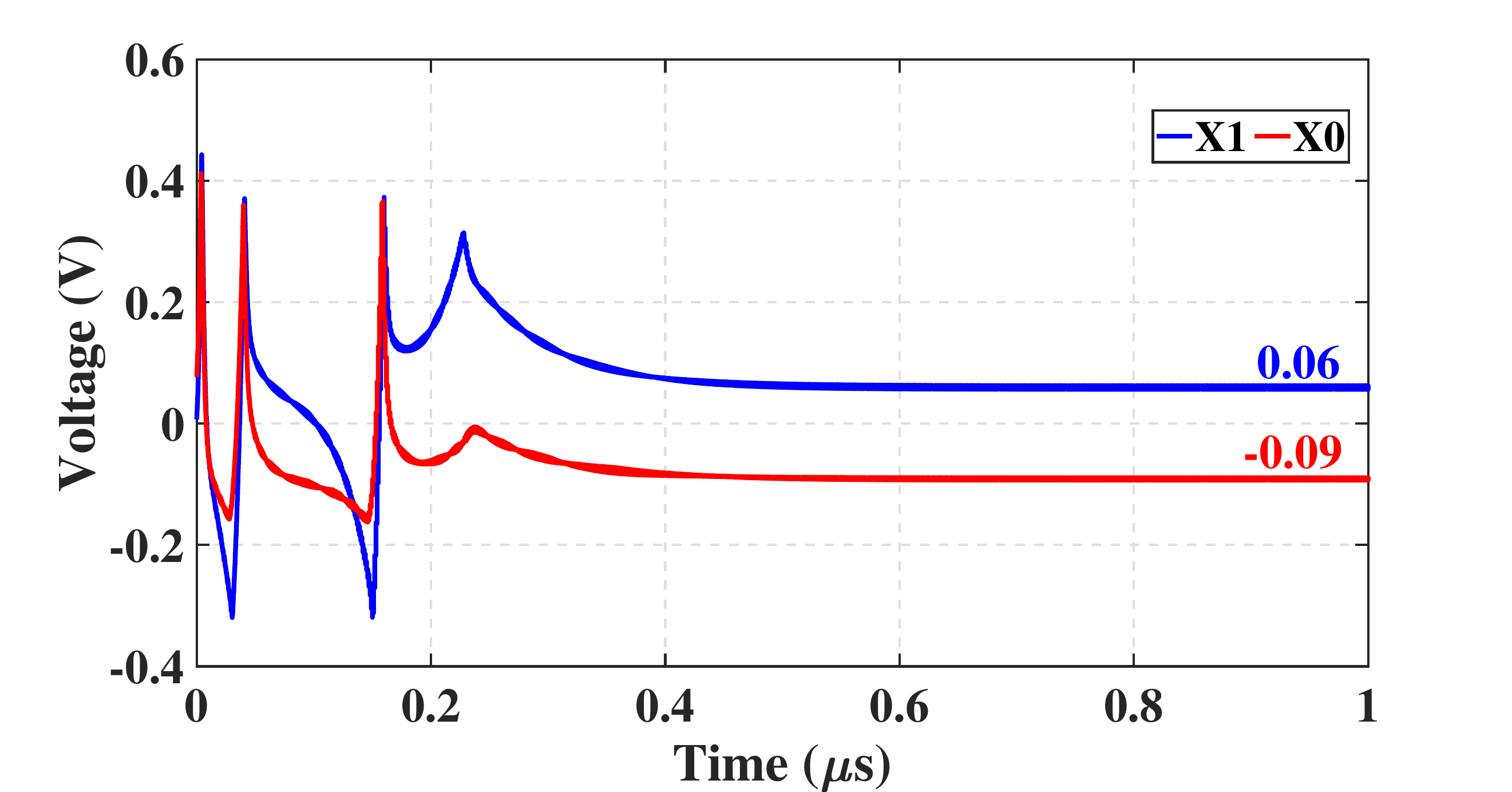}
\caption{The output of the system with a 2 × 2 mixed matrix values.}
\vspace{-2em}
\label{fig:2mixed}
\end{figure}

For a system with an input signal in the form of a 2 × 1 matrix, the output values can converge within 200 ns when the implemented matrix consists entirely of positive or negative values. However, when the implemented matrix contains both positive and negative values, the output values converge within 400 ns. 

Furthermore, if we extend the structure of Fig. 4 to implement an 8 × 8 matrix, where the elements of the matrix can be positive or negative, both the input and output will also be 8 × 1 matrices. The matrix we input is as follows:
\begin{equation}
B = 
{\small
\begin{bmatrix}
0.375 & 0.225 & 0.225 & 0.225 & 0.225 & 0.225 & 0.225 & 0.225
\end{bmatrix}^{\mathrm{T}}},
\label{eq:bt}
\end{equation}
where the first element is 0.375, while the other elements are 0.225. And the 8 × 8 matrix implemented by the expanded system is:
\begin{equation}
A =
\begin{bmatrix}
0.5  & -1  & -1  & -1  & -1  & -1  & -1  & -1 \\
-1  &  0.5  & -1  & -1  & -1  & -1  & -1  & -1   \\
-1  & -1  &  0.5  & -1  & -1  & -1  & -1  & -1   \\
-1  & -1  & -1  &  0.5  & -1  & -1  & -1  & -1   \\
-1  & -1  & -1  & -1  &  0.5  & -1  & -1  & -1   \\
-1  & -1  & -1  & -1  & -1  &  0.5 & -1  & -1   \\
-1  & -1  & -1  & -1  & -1  & -1  &  0.5  & -1   \\
-1  & -1  & -1  & -1  & -1  & -1  & -1  & 0.5
\end{bmatrix},
\label{eq:at}
\end{equation}
the results obtained from the system simulation, where the $K_{vco}$ of the oscillator is 300 MHz, are presented in Fig. \ref{fig:300m}, with the following results:
\begin{equation}
{\small
X=
\begin{bmatrix}
0.05 & -0.05 & -0.05 & -0.05 & -0.05 & -0.05 & -0.05 & -0.05
\end{bmatrix}^{\mathrm{T}}}.
\label{eq:xt}
\end{equation}
\vspace{-2em}

\begin{figure}[ht]
\centering
\includegraphics[scale=0.2]{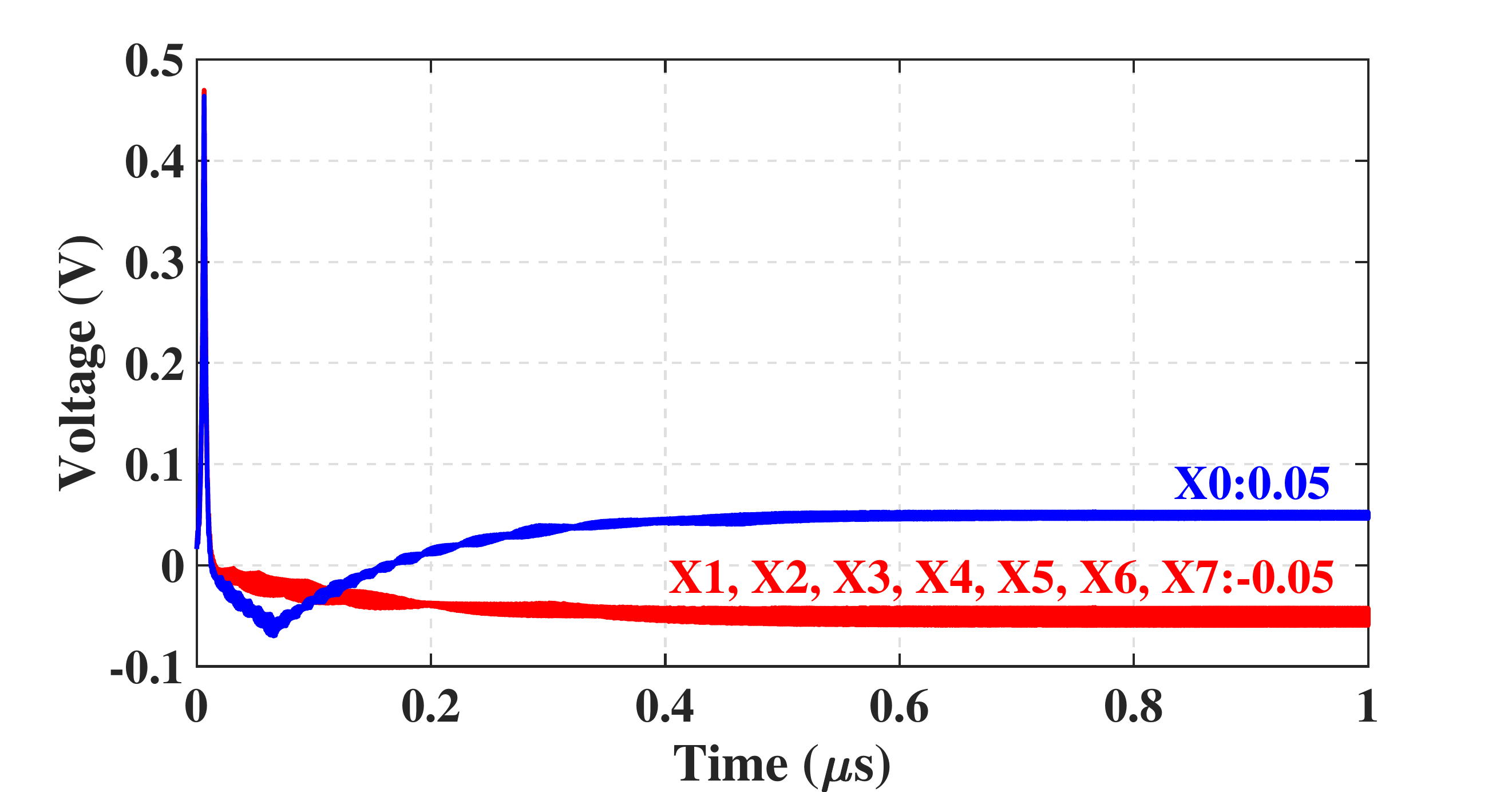}
\caption{The output of the system having 8 × 8 mixed values with $K_{vco}$ of $300$ MHz.}
\label{fig:300m}
\end{figure}

\begin{table*}[ht]
\caption{The comparison of this work and other computing systems}
\label{tab:3}
\centering
\renewcommand{\arraystretch}{1.2}
\setlength{\tabcolsep}{8pt}
\resizebox{\textwidth}{!}{%
\begin{tabular}{|c|c|c|c|c|c|c|c|} 
\hline 
& \cite{b36}& \cite{b37}&\cite{b35}& \cite{b38}& \cite{b39}&  \cite{LinBarrowsCarevelli2024}&This work\\ 
\hline 
Computing method& digital method& digital method &analog method& analog method& analog method&analog method& analog method\\ 
\hline 
Matrix size ($A$)& 24 × 24&  24 × 24 & 10 × 10& 2 × 2& 3 × 3& 10 × 10&8 × 8\\ 
\hline
 Computation time ($\mu$s)& 136&  551 &12.08& 80& 4000 & 4E5, $O(n^2)$& 
10\\\hline
 Resource usage (Area)& very large& very large &small& small& small&very small&very small\\\hline
 MOPs/s& 317& 16.2& N/A& N/A& N/A& N/A&34.1\\\hline
 GOPs/W& 0.27& 0.03& 1025 ×  CPU& N/A& N/A& N/A&5.7\\\hline
Power (mW)& 1150$\dag$\tnote{a} & 536$\ddag$\tnote{b} & N/A& N/A& N/A& N/A&6\\\hline
 Energy ($\mu$J)& 156.4& 295.33&N/A &N/A &N/A &N/A &0.06\\\hline
\end{tabular}%
}
\begin{tablenotes}[flushleft]
\item[a] $\dag$$P_{\text{dynamic}} = k \cdot N_{\text{LUT}} \cdot f$, where: $k = {0.25} ~\mu W/LUT/MHz$ (power constant for Spartan-3), $N_{\text{LUT}} = 24^2 \times 222 = 127,\!872$, $f = {40}{~MHz}$ (operating frequency), $P_{\text{static}} = 500~mW$. \textit{Source:} Xilinx Spartan-3 datasheet (DS099).
\item[b] $\ddag$$P_{\text{dynamic}} = k \cdot N_{\text{LUT}} \cdot f$, where: $k = {0.2}{~\mu W/ LUT/MHz}$ (power constant), $N_{\text{LUT}} = 29,500$ (5 Nios cores × 5,900 LUTs/core), $f = {40}{~MHz}$ (operating frequency), $P_{\text{static}} = {300}{~mW}$ (FPGA leakage power). \textit{Source:} Altera EP20K1500E power models.
\end{tablenotes}
\end{table*}

\begin{figure}[H]
\centering
\includegraphics[scale=0.2]{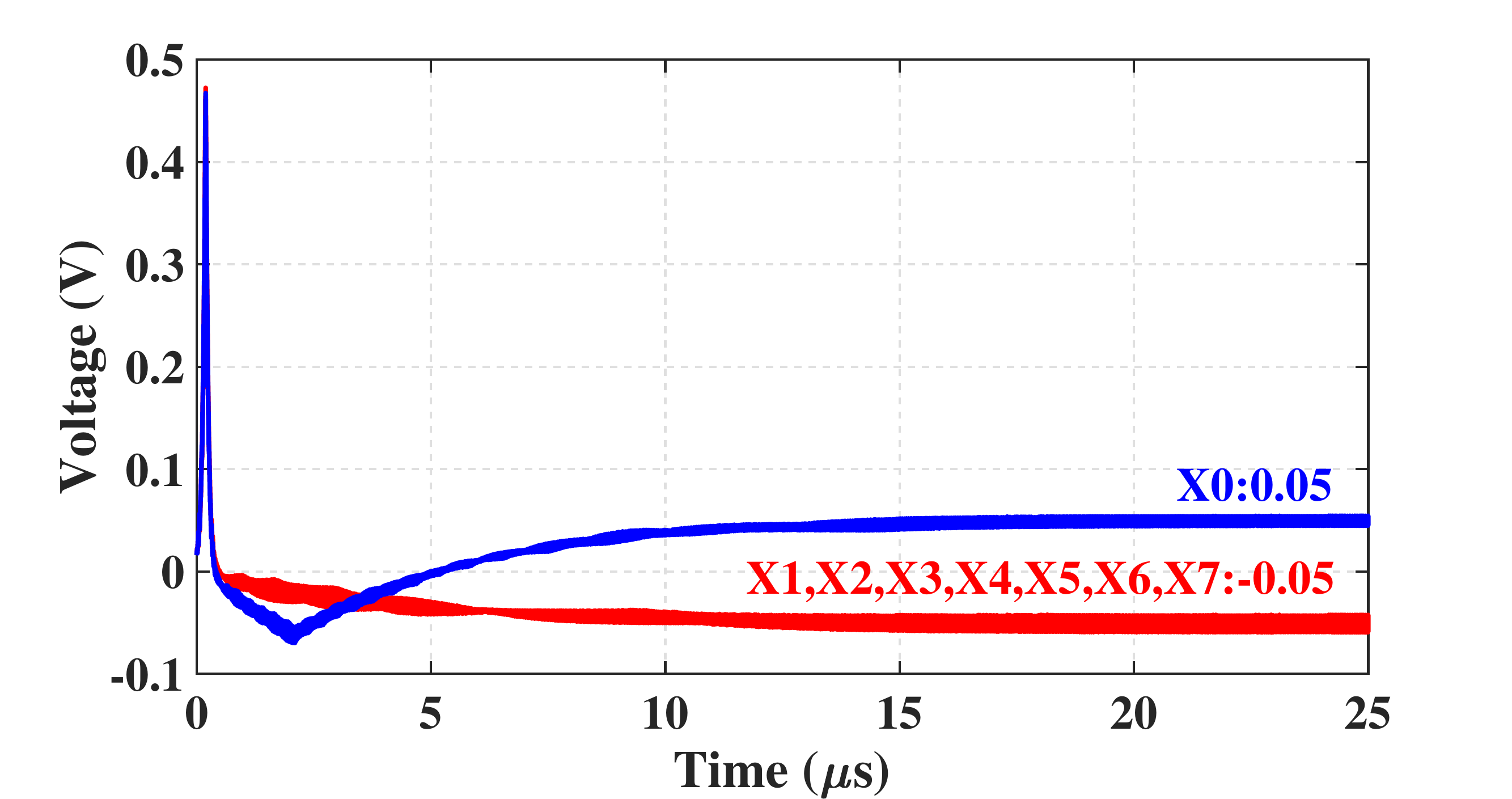}
\caption{The output of the system having 8 × 8 mixed values with $K_{vco}$ of $20$ MHz.}
\label{fig:20mhz}
\end{figure}

As shown in Fig. \ref{fig:300m}, for the computation of the 8 × 8 linear equation, the proposed system delivers the results within 400 ns when $K_{vco}$ of the oscillator is 300 MHz. Compared to computations using digital circuit chips or FPGAs, this computation time exhibits a significant advantage, the resources consumed are significantly reduced at the same time. We will provide a comparative discussion in the next subsection.

There is a trade-off between power and $K_{vco}$ of the oscillator. A reduction in power consumption inherently leads to a decrease in $K_{vco}$. As illustrated in Fig. \ref{fig:20mhz}, to enable low-power operation, $K_{vco}$ is reduced to 20 MHz. Consequently, the converging time extends to 10~$\mu$s.

\subsection{Comparison}
In this part, we will give performance analyses and a comparison of the proposed architecture with existing architectures. This comparison includes both digital and analog computation approaches. The digital computation method means traditional digital approaches, such as digital circuit chips and FPGA-based solutions. Analog computation focuses on analog computing circuits, which have received significant research attention in recent years \cite{b35}. The key performance comparison between the proposed scheme and other digital and analog computing systems is presented in Table \ref{tab:3}. From this table, it is clear that, compared with other similar circuit structures, the work presented in this paper features a relatively simple design with lower resource usage and faster computation time. 

In addition, it operates efficiently at a voltage as low as 1 V, resulting in low power consumption, based on the previous circuit design and analysis, the energy efficiency $\eta$ of the 8 × 8 computing system can be evaluated as follows: 
\begin{equation}
\eta = \frac{N}{P*T}  ,  
\label{eq:energy}
\end{equation}
where $N$ denotes the operations (OPs) of the system within the converging time, $P$ represents the power of the whole system and $T$ is the converging time of the circuit. Assuming that the OPs in the proposed 8 × 8 system is equivalent to that of the LU factorization, which is $\frac{2}{3} ~$×$~ 8^3$, the resulting value of $N$ is 341. $P$ is determined by the number of integrators utilized in the system. Assuming that each integrator consumes approximately 0.15~mW and a total of 40 integrators are deployed, the total power consumption is estimated to be around 6 mW. The energy efficiency of the 8 × 8 system is calculated to be 5.7 GOPs/W, where GOPs means giga operations per second, as determined by (36). 


\section{Conclusion}
This paper proposes a ring oscillator based linear algebra system of equations to overcome the limitations of analog computing in terms of chip area, power consumption, and computation time. This system is capable of solving the linear equation $AX=B$, where $B$ is the input matrix and $X$ is the output. The solving process has been analyzed and verified when the matrix realized by the system is 2 × 2 or 8 × 8 matrix, and the elements of the matrix can be positive or negative. The calculation can be completed in 0.4 \textmu s for the system with a 8 × 8 matrix when the $K_{vco}$ of the oscillator is 300 MHz. Due to its small area and low power consumption, this architecture is expected to be scalable for solving 16 × 16 or 32 × 32 linear algebraic matrix equations, delivering outstanding performance in terms of computational speed. As the demand for matrix computations continues to grow, this kind of circuit structure has great potential for future efficient matrix computations.
\vspace{-1em}
\section*{Appendix}
 For practical applications, since the values of the matrix $A$ are generated randomly, it can produce excessively large or small values, causing the output $X$ to exceed the limits of the circuit. To mitigate this issue, for a linear equation of $AX=B$, due to circuit limitations, it is assumed that all elements of the matrix $B$ are restricted within the range of [-0.5, 0.5]. The objective is to scale the elements of the matrix $A$ so that all elements of the matrix $X$ are restricted within the range of [-0.5, 0.5]. According to the matrix norm inequality, the following relation holds:
\begin{equation}
\|X\| \leq \|A^{-1}\| \cdot \|B\|.
\label{eq:xnorm}
\end{equation}
In the case of the infinity norm,\footnote{The infinity norm (or max-norm) of a matrix is defined as the maximum absolute row sum. This is obtained by summing the absolute values of the elements in each row and then taking the maximum of these sums.} this inequality can be expressed as:
\begin{equation}
\|X\|_{\infty} \leq \|A^{-1}\|_{\infty} \cdot \|B\|_{\infty},
\label{eq:xinfi}
\end{equation}
which further implies:
\begin{equation}
\max(\lvert X \rvert) \leq \| A^{-1} \|_{\infty} \cdot \max(\lvert B \rvert),
\label{eq:xmax}
\end{equation}
where $\max(\lvert X \rvert)$ and $\max(\lvert B\rvert)$ denote the maximum absolute values of the entries in matrices $X$ and $B$, respectively. Let $A_{\text{scaled}}$ denote the scaled version of matrix $A$, defined as:
\begin{equation}
A_{\text{scaled}} = A \cdot \text{factor}_{\text{scale}},
\label{eq:ascaled}
\end{equation}
where $\text{factor}_{\text{scale}}$ is the scaling factor. Consequently, the inverse of the scaled matrix is given by:
\begin{equation}
A_{\text{scaled}}^{-1} = \frac{A^{-1}}{\text{factor}_{\text{scale}}}
\label{eq:ascaled_2}
\end{equation}
Substituting this into \eqref{eq:xmax}, we obtain:
\begin{equation}
\max(\lvert X \rvert) \leq \|A_{\text{scaled}}^{-1} \|_{\infty} \cdot 0.5.
\label{eq:xmax_scale}
\end{equation}
It follows that the scaling factor must satisfy:
\begin{equation}
\text{factor}_{\text{scale}} \geq \| A^{-1} \|_{\infty}.
\label{eq:factor}
\end{equation}
From the definition of the condition number in the infinity norm\cite{trefethen1997numerical}:
\begin{equation}
\kappa_{\infty}(A)=\|A^{-1}\|_{\infty} \cdot \|A\|_{\infty},
\label{eq:subinquality}
\end{equation}
where $\kappa_{\infty}(A)$ represents the condition number of $A$ in the infinity norm. Substituting \eqref{eq:subinquality} into \eqref{eq:factor} yields:
\begin{equation}
\text{factor}_{\text{scale}} \geq \frac{\kappa_{\infty}(A)}{\|A\|_{\infty}}.
\label{eq:factor_final}
\end{equation}
This condition ensures that any matrix can be accurately represented in the system if the scaling factor is chosen according to the inequality above. For general nonsingular matrices, a conservative estimate for the infinity-norm condition number is typically assumed as:
\begin{equation}
\kappa_\infty(A) \lesssim 10^2 \sim 10^3,
\label{eq:kappa}
\end{equation}
which implies an upper bound on the inverse norm:
\begin{equation}
\|A^{-1}\|_\infty \lesssim \frac{10^3}{\|A\|_\infty}.
\label{eq:kappa_with_A}
\end{equation}
This estimate is widely utilized in numerical analysis and circuit design to mitigate overflow risks and maintain bounded outputs when solving the linear system $AX = B$, particularly under the constraint $\max(|B|) \leq 0.5$~\cite{trefethen1997numerical,higham2002accuracy,golub2013matrix}. The computational complexity of calculating the infinity norm of a matrix is $\mathcal{O}(n^2)$. Therefore, the overall complexity of the scaling operation remains $\mathcal{O}(n^2)$.

\end{document}